\documentclass[usegraphicx]{mn2e}

\usepackage{amssymb}
\usepackage{mathptmx}
\usepackage{url}

\newcommand{\Mpc}{\rm\thinspace Mpc}
\newcommand{\kpc}{\rm\thinspace kpc}

\newcommand{\km}{\rm\thinspace km}

\newcommand{\yr}{\rm\thinspace yr}

\newcommand{\s}{\rm\thinspace s}

\newcommand{\Msun}{\hbox{$\rm\thinspace M_{\odot}$}}

\newcommand{\Msunpyr}{\hbox{$\Msun\yr^{-1}\,$}}

\newcommand{\keV}{\rm\thinspace keV}

\newcommand{\erg}{\rm\thinspace erg}

\newcommand{\ergps}{\hbox{$\erg\s^{-1}\,$}}

\newcommand{\kmps}{\hbox{$\km\s^{-1}\,$}}

\newcommand{\kmpspMpc}{\hbox{$\kmps\Mpc^{-1}$}}

\newcommand{\Lsun}{\hbox{$\rm\thinspace L_{\odot}$}}

\newcommand{\Zsun}{\hbox{$\thinspace \mathrm{Z}_{\odot}$}}

\begin{document}
\title{Enrichment in the Centaurus cluster of galaxies}

\author[J.S. Sanders and A.C. Fabian]{J.S.  Sanders\thanks{E-mail:
    jss@ast.cam.ac.uk} and
  A.C. Fabian\\
  Institute of Astronomy, Madingley Road, Cambridge. CB3 0HA}
\maketitle

\begin{abstract}
  We perform a detailed spatially-resolved, spectroscopic, analysis of
  the core of the Centaurus cluster of galaxies using a deep
  \emph{Chandra} X-ray observation and \emph{XMM-Newton} data. The
  Centaurus cluster core has particularly high metallicity, up to
  twice Solar values, and we measure the abundances of Fe, O, Ne, Mg,
  Si, S, Ar, Ca and Ni. We map the distribution of these elements in
  many spatial regions, and create radial profiles to the east and
  west of the centre. The ratios of the most robustly determined
  elements to iron are consistent with Solar ratios, indicating that
  there has been enrichment by both Type Ia and Type II supernovae.
  For a normal initial stellar mass function it represents the
  products of about $4\times 10^{10}\Msun$ of star formation. This
  star formation can have occured either continuously at a rate of
  $5\Msunpyr$ for the past 8~Gyr or more, or was part of the formation
  of the central galaxy at earlier times. Either conclusion requires
  that the inner core of the Centaurus cluster has not suffered a
  major disruption within the past 8~Gyr, or even longer.
\end{abstract}

\begin{keywords}
  X-rays: galaxies --- galaxies: clusters: individual: Centaurus ---
  intergalactic medium --- cooling flows
\end{keywords}

\section{Introduction}
X-ray observations of the intracluster medium in clusters of galaxies
allows us to measure the integrated enrichment in these objects over
their lifetime. The two main contributors to the metals in clusters
are Type~Ia and Type~II supernovae. The expected ratio of the yields
of various elements are quite different. Type~II explosions should
produce significantly more $\alpha$-element yield relative to Fe than
Type~Ia events.  Therefore by studying the relative abundances of
elements, it is possible to determine the relative contribution of
different enrichment mechanisms, and so learn about the star formation
history of these objects (Renzini et al 2004).

Local clusters of galaxies are enriched to around 1/3 of the solar
value (as determined mainly by the Fe abundance; e.g. Edge \& Stewart
1991). Clusters with highly peaked surface brightness distributions,
traditionally known as cooling flow clusters, have higher average
X-ray emission-weighted metallicities than those without (Allen \&
Fabian 1998). In addition these peaked clusters show metallicity
gradients, with large abundances in their cores (e.g. Fukazawa et al
1994; De Grandi \& Molendi 2001).

Data from the \emph{ASCA} X-ray observatory allowed many studies to be
made of the metallicity of clusters, given its better spectral
resolution compared to previous missions. Mushotzky et al (1996)
examined four clusters in the temperature range 2.5-5~keV, removing
the centrally peaked regions. They concluded the abundance ratios
observed were consistent with Type~II supernovae. Fukazawa et al
(1998) examined 40 clusters, concluding that relative contribution of
Type~II enrichment increases towards richer systems, with Type~Ia
enrichment being more important in lower mass systems.

Recently Baumgartner et al (2005) examined all the cluster
observations in the \emph{ASCA} archive, performing a stacked
analysis. Their conclusion is that generally Si and Ni are
overabundant with respect to Fe, but Ar and Ca are very underabundant.
Therefore the degree of enrichment of $\alpha$-elements is not the
same. The Fe, S and S abundances did not consistently match a
combination of theoretical Type~Ia and Type~II products.

Recently \emph{XMM-Newton} and \emph{Chandra} have been used to make
detailed analyses of clusters, examining the variation of different
metals as a function of radius, and mapping (Sanders et al 2004).
Matsushita, Finoguenov \& B\"ohringer (2003) examined an \emph{XMM}
observation of M87. They found no evidence for a gradient in Fe/Si,
but found O to have a flatter gradient than Fe or Si. From the lack of
O emission in the very central regions they concluded that Type~Ia
enrichment is dominant in there. Werner et al (2006) examined the
2A~0335+096 cluster in detail, finding a contribution to the
enrichment of around 75~per~cent from Type~Ia supernovae, and
25~per~cent Type~II supernovae.

The Centaurus cluster, Abell 3526, is a nearby ($z=0.0104$; Lucey,
Currie \& Dickens 1986) bright galaxy cluster ($L_{X,2-10\keV} = 2.9
\times 10^{45} \ergps$; Edge et al 1990).

The cluster is notable for its steep abundance gradient (Fukazawa et
al 1994; Ikebe et al 1999; Allen et al 2001). The central region is
very rich in metals (Sanders \& Fabian 2002), reaching twice solar
abundance in iron (Fabian et al 2005). In addition the high
metallicity region is mostly to the west of the cluster (Sanders \&
Fabian 2002), and there appears to be a drop in abundance in the very
central region. This drop is not due to resonance scattering (Sanders
\& Fabian 2006).

Any abundances we show in this paper are relative to the commonly used
photospheric solar metallicities of Anders \& Grevesse (1989). In this
set of abundances Fe has a number density of $4.68 \times 10^{-5}$
relative to H. We use a redshift for the Centaurus cluster of 0.0104,
and assume $H_0 = 70 \kmpspMpc$, which gives a scale of 213~pc per
arcsec. All uncertainties shown are 1-$\sigma$. Positions use J2000
coordinates.

\section{Data preparation}
\subsection{\emph{Chandra} datasets}
\begin{table}
  \caption{\emph{Chandra} datasets examined in this paper and their
    observation date. The exposure shown is the
    total exposure after periods with the flares are removed. The
    combined exposure is 195.4~ks. The offsets applied to the
    observations are in the $y$-coordinate of the CCD.}
  \label{tab:datasets}
  \begin{tabular}{lllll}
    \hline
    Obs  & Sequence & Date       & $y$-offset (arcmin) &  Exposure (ks) \\ \hline
    504  & 800012   & 2000-05-22 & -1.0              & 21.5 \\
    505  & 800013   & 2000-06-08 & -1.0              & 9.7  \\
    4954 & 800401   & 2004-04-01 & -1.3              & 80.0 \\
    4955 & 800401   & 2004-04-02 & -0.8              & 40.1 \\
    5310 & 800401   & 2004-04-04 & -1.8              & 44.1 \\
    \hline
  \end{tabular}
\end{table}

The \emph{Chandra} datasets examined in this paper are listed in
Table~\ref{tab:datasets}. The ACIS-S3 aimpoint was offset from the
X-ray peak slightly by different amounts to dither the observation to
help remove CCD artifacts.

We investigated the results using both the standard CALDB calibration,
and using data corrected for charge transfer inefficiency (CTI) using
the Penn State University corrector (PSU; Townsley et al 2002a,
2002b). CTI has degraded the energy resolution of the ACIS CCDs since
the launch of \emph{Chandra}. ACIS contains both front-illuminated
(FI) and back-illuminated (BI) CCDs. The effect of CTI has been
greatest on the FI CCDs. It is thought the damage is due to focused
protons. The gates of the FI CCDs are unprotected by covering silicon,
and are damaged more quickly.

The standard analysis tools correct data from the FI CCDs for this
effect, but not for BI data. The ACIS-S3 CCD is a BI CCD, and is a
popular choice of primary detector for \emph{Chandra} users due to its
comparatively high effective area at lower energies, and its higher
spectral resolution. The degradation of spectral resolution changes
the width of spectral lines, and so is potentially important for
metallicity determination. We therefore investigated whether
correcting for CTI has a significant effect in abundance
determinations, which we discuss below.

For the non-CTI corrected data, we reprocessed each dataset using the
most recent appropriate gain file, acisD2000-01-29gain\_ctiN0003.fits,
and applying time-dependent gain correction. Using a lightcurve from
the ACIS-S5 CCD (which is a back-illuminated CCD, like ACIS-S3) in the
2.5 to 7~keV band (as recommended by Markevitch 2002), we filtered
each event file using the \textsc{ciao} \textsc{lc\_clean} tool. This
yielded a total exposure time of 195.4~ks.

When CTI correcting, we reprocessed each level-1 event file with the
PSU corrector, removed standard event grades, and applied the same
time filtering as above.

We used blank-sky background observations when fitting the spectra
from each of the datasets. In order for the background spectra to be
independent for each dataset, we split a single blank sky background
up into three for the 4954, 4955 and 5310 observations, keeping the
ratio of the background exposure to the observation to be the same. As
the \emph{Chandra} ACIS background is slowly increasing as a function
of time, we therefore adjusted the exposure of each of the blank sky
event files in order for the 9 to 12~keV rate of photons in the
background observation on the S3 CCD to be the same as the
observation.

\subsection{Creating spectra and responses}
When spectral fitting the \emph{Chandra} spectra, rather than attempt
simultaneously fitting of spectra from each of the datasets, we added
together the spectra before fitting. In addition we generated response
and ancillary response files for each spectral region for each of the
datasets. The response files for the datasets were added together,
weighting according to the number of photons between 0.5 and 7~keV in
the corresponding spectrum. The background spectra appropriate for
each foreground dataset were merged together to make a total
background spectrum.  This was done by discarding random photons from
the background spectra, effectively making the exposure time shorter,
until the ratio of the exposure time of each background spectrum to
the total background matched the ratio of the exposure times of each
respective foreground spectrum to the total foreground. We only
considered the ACIS-S3 when spectral fitting.

The adding procedure is valid of the response of the detector is
similar between each observation. This is the case, except for the
build up of contaminant on the ACIS detector over time. We tested the
method by comparing the results from spectral fitting against those
from fitting the spectra simultaneously. We saw no systematic changes
in results.

For the non-CTI corrected data we generated response matrices and
ancillary response matrices using the \textsc{mkacisrmf} and
\textsc{mkwarf} tools, weighting each CCD region according to the
number of counts between 0.5 and 7~keV. We used a single response
matrix across the entire ACIS-S3 CCD for the CTI corrected data,
generated with high energy resolution. \textsc{mkwarf} was applied to
generate ancillary responses, but using the QEU file of the PSU
corrector.

\subsection{\emph{XMM-Newton} observation}
In Section~\ref{sect:profiles} we examine the radial variation of the
properties of the cluster. In this analysis we also examine an
\emph{XMM-Newton} observation of the cluster which goes out to larger
radius than the \emph{Chandra} observations. We examined \emph{XMM}
observation identifier 0046340101. The data were reduced using the
standard tools, and were filtered using the same criteria as used by
Read \& Ponman (2003) to create their backgrounds. This yielded
26.7~ks and 31.4~ks of time on the MOS1 and MOS2. There was no PN data
remaining after filtering. Background spectra were taken from the
blank-sky datasets of Read \& Ponman (2003), normalising to match the
count rate of detector outside of the telescope field of view. The
exposure times of the backgrounds were adjusted so that the count rate
in the 9 to 11~keV band, where there is little cluster signal, matched
the foreground spectra.

When spectral fitting, the two MOS datasets were fit simultaneously,
but the relative normalisations of the two instruments were allowed to
vary to account for any variation in effective area. The spectra were
fit between 0.5 and 7~keV. Except for the simultaneous fit of the two
datasets, we used the same spectral fitting procedure as the
\emph{Chandra} data, which is detailed below.

\section{Two dimensional analysis}
We selected regions on the surface brightness images containing a
signal to noise ratio of 200 or greater (over 40\,000 counts) using
the contour binning technique (Sanders 2006). The algorithm selects
regions which have the same surface brightness on an image by binning
pixels with nearest fluxes on the smoothed map until a signal to noise
threshold is reached. In this case the image was smoothed using the
accumulative smoothing algorithm to give a minimum signal to noise of
30. We also constrained the geometry of the bins so that new pixels
were not added if they are further away than 1.8 times the radius of a
circle with the same area as the bin currently.  This produces bins
which are more compact than otherwise would be the case.

\begin{table*}
  \caption{Point sources excluded from the \emph{Chandra} data, in RA order. Also shown is the
    radius of the excluded circle in arcsec.}

  \begin{minipage}[t]{0.33\textwidth}
  \begin{tabular}{lll}
    \hline
    RA & Dec & Radius ($''$) \\ \hline
       $12^{h}48^{m}22.61^{s}$ & $-41^\circ 21' \: 36.1''$ &        4.9 \\
       $12^{h}48^{m}23.53^{s}$ & $-41^\circ 21' \: 27.1''$ &        4.4 \\
       $12^{h}48^{m}26.14^{s}$ & $-41^\circ 17' \: 56.3''$ &        5.4 \\
       $12^{h}48^{m}29.56^{s}$ & $-41^\circ 18' \: 49.0''$ &        2.8 \\
       $12^{h}48^{m}29.84^{s}$ & $-41^\circ 22' \: 49.3''$ &        3.7 \\
       $12^{h}48^{m}30.89^{s}$ & $-41^\circ 19' \: 31.3''$ &        6.1 \\
       $12^{h}48^{m}30.93^{s}$ & $-41^\circ 22' \: 43.2''$ &        2.8 \\
       $12^{h}48^{m}31.33^{s}$ & $-41^\circ 18' \: 26.4''$ &        1.7 \\
       $12^{h}48^{m}32.51^{s}$ & $-41^\circ 18' \: 32.1''$ &        2.9 \\
       $12^{h}48^{m}35.54^{s}$ & $-41^\circ 15' \: 02.5''$ &        3.4 \\
       $12^{h}48^{m}36.51^{s}$ & $-41^\circ 18' \: 21.3''$ &        2.4 \\
       $12^{h}48^{m}38.99^{s}$ & $-41^\circ 19' \: 41.0''$ &        2.2 \\
       $12^{h}48^{m}39.26^{s}$ & $-41^\circ 17' \: 39.0''$ &        2.8 \\
       \hline
       \end{tabular}
  \end{minipage}
  \begin{minipage}[t]{0.33\textwidth}
  \begin{tabular}{lll}
    \hline
    RA & Dec & Radius ($''$) \\ \hline
       $12^{h}48^{m}39.29^{s}$ & $-41^\circ 20' \: 24.6''$ &        2.5 \\
       $12^{h}48^{m}39.75^{s}$ & $-41^\circ 19' \: 27.0''$ &        1.5 \\
       $12^{h}48^{m}41.45^{s}$ & $-41^\circ 20' \: 48.0''$ &        1.9 \\
       $12^{h}48^{m}41.63^{s}$ & $-41^\circ 19' \: 28.0''$ &        2.3 \\
       $12^{h}48^{m}42.40^{s}$ & $-41^\circ 17' \: 11.5''$ &        2.1 \\
       $12^{h}48^{m}42.84^{s}$ & $-41^\circ 16' \: 09.5''$ &        2.8 \\
       $12^{h}48^{m}42.99^{s}$ & $-41^\circ 22' \: 48.8''$ &        4.4 \\
       $12^{h}48^{m}43.04^{s}$ & $-41^\circ 15' \: 41.7''$ &        3.5 \\
       $12^{h}48^{m}43.53^{s}$ & $-41^\circ 20' \: 06.4''$ &        1.9 \\
       $12^{h}48^{m}44.17^{s}$ & $-41^\circ 16' \: 10.5''$ &        1.9 \\
       $12^{h}48^{m}46.62^{s}$ & $-41^\circ 15' \: 39.5''$ &        2.0 \\
       $12^{h}48^{m}48.76^{s}$ & $-41^\circ 16' \: 07.1''$ &        1.8 \\
       $12^{h}48^{m}53.44^{s}$ & $-41^\circ 19' \: 12.9''$ &        1.2 \\
       \hline
       \end{tabular}
  \end{minipage}
  \begin{minipage}[t]{0.33\textwidth}
  \begin{tabular}{lll}
    \hline
    RA & Dec & Radius ($''$) \\ \hline
       $12^{h}48^{m}55.32^{s}$ & $-41^\circ 19' \: 02.2''$ &        1.0 \\
       $12^{h}48^{m}56.17^{s}$ & $-41^\circ 19' \: 23.8''$ &        0.8 \\
       $12^{h}48^{m}56.90^{s}$ & $-41^\circ 14' \: 50.1''$ &        3.1 \\
       $12^{h}48^{m}56.96^{s}$ & $-41^\circ 19' \: 21.2''$ &        1.3 \\
       $12^{h}48^{m}59.82^{s}$ & $-41^\circ 15' \: 59.0''$ &        2.4 \\
       $12^{h}48^{m}59.99^{s}$ & $-41^\circ 14' \: 40.7''$ &        4.3 \\
       $12^{h}49^{m}01.57^{s}$ & $-41^\circ 20' \: 34.5''$ &        1.5 \\
       $12^{h}49^{m}02.50^{s}$ & $-41^\circ 14' \: 41.2''$ &        4.3 \\
       $12^{h}49^{m}03.16^{s}$ & $-41^\circ 18' \: 49.9''$ &        1.2 \\
       $12^{h}49^{m}03.62^{s}$ & $-41^\circ 20' \: 50.2''$ &        1.8 \\
       $12^{h}49^{m}05.84^{s}$ & $-41^\circ 15' \: 39.8''$ &        2.4 \\
       $12^{h}49^{m}06.24^{s}$ & $-41^\circ 17' \: 51.1''$ &        2.5 \\
       \\
       \hline
  \end{tabular}
  \end{minipage}
  \label{table:excludedpts}
\end{table*}

In all the \emph{Chandra} analyses, point sources listed in
Table~\ref{table:excludedpts} were excluded from the regions. These
point sources are not shown in the abundance maps.

We extracted spectra from each of the regions, group each spectrum to
contain a minimum of 20 counts per spectral channel. We fitted the
spectra using spectral models made up of one, two and three
temperature components. In the fits the O, Ne, Mg, Si, S, Ar, Ca, Fe
and Ni abundances were allowed to be free. We assumed each temperature
component had the same abundance. In the fits we also allowed the
temperatures, normalisations of the components and Galactic absorption
to be free. The spectra were fit between 0.5 and 7~keV to minimise the
$\chi^2$ statistic. We used an F-test to decide between the single,
double and triple component fits. If the double temperature fit gave
an improvement over the single temperature fit with a probability of
0.27~per~cent improvement by chance (corresponding to 3-$\sigma$),
then we used that.  If the triple component fit gave the same minimum
improvement over the double fit, then that was used.

\subsection{Spectral codes}
\label{sect:speccodes}
There are a few spectral models in use in X-ray astronomy for
optically thin, collisional plasmas as found in clusters. These
include \textsc{mekal} (Mewe, Gronenschild \& van den Oord 1985; Mewe,
Lemen \& van den Oord 1986; Kaastra 1992; Liedahl, Osterheld \&
Goldstein 1995) in \textsc{xspec} using the ionisation balance of
Arnaud \& Rothenflug 1985, Arnaud \& Raymond 1992. \textsc{xspec}
(Arnaud 1996) is the standard X-ray spectral fitting package, which we
use here. \textsc{mekal} can be calculated in real-time which spectral
fitting, or used by interpolation in temperature of precalculated
spectra. We use the calculation option as the tables are inaccurate at
low temperatures, and have poor spectral resolution.

A more recent code is \textsc{apec} (Smith et al 2001b), also
available in \textsc{xspec}, where we use the latest available
version, 1.3.1.  This model uses precalculated tables generated from
the Astrophysical Plasma Emission Database (APED; Smith et al 2001a).
The tables are more accurate than the \textsc{xspec} \textsc{mekal}
tables, as the spectral lines are stored as separate list to the
continuua spectra, and so spectral resolution is not an issue. It has
also enabled us to include the effects of line width in the model (now
included in the version provided in \textsc{xspec}).

The \textsc{mekal} code has also been developed over time, but these
changes are not present in the version in \textsc{xspec}. They are,
however, present in the spectral analysis package \textsc{spex}
(Kaastra 2000). Until now, it has not been possible to use the updated
version of \textsc{mekal}, which we call \textsc{spex} in this paper
for convenience, in \textsc{xspec}. Nevertheless, it is easy to
generate continua spectra and lists of lines for the \textsc{spex}
model in the \textsc{spex} spectral fitting package for a set of
temperatures. We were able to take these data and generate
\textsc{apec} format table models from it, again storing the continuum
from separately from the lines. This has allowed us to compare the
results using the three different models with the same analysis
techniques in the same analysis package, \textsc{xspec}. Reassuringly,
the results obtained on our CCD resolution spectra using
\textsc{mekal} and \textsc{spex} are very similar.

\begin{figure}
  \includegraphics[width=\columnwidth]{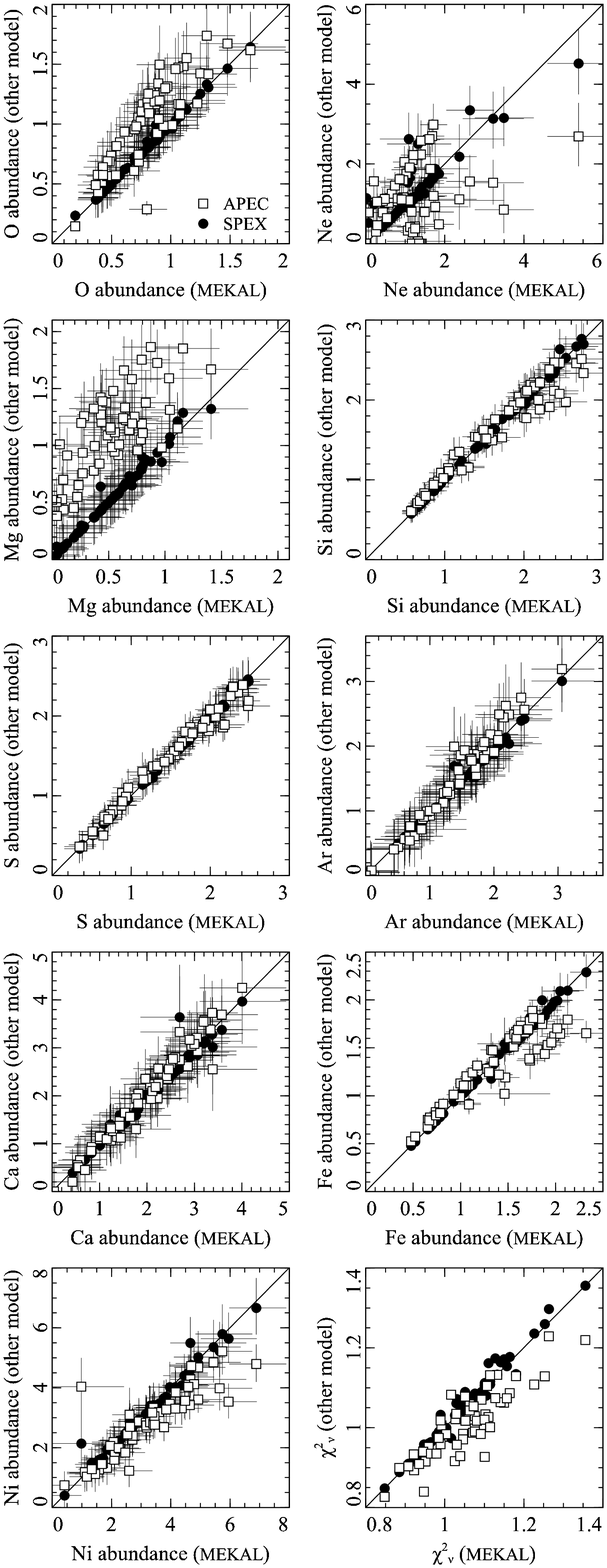}
  \caption{Comparison of metallicities and reduced $\chi^2$ obtained
    with the \textsc{apec} and \textsc{spex} spectral codes against
    those using \textsc{mekal}. As \textsc{spex} is based on
    \textsc{mekal}, the results obtained are very similar.}
  \label{fig:codecompar}
\end{figure}

To examine the differences more systematically we plot in
Fig.~\ref{fig:codecompar} the abundances obtained from the
\textsc{apec} and \textsc{spex} codes for each of the regions against
the metallicity using \textsc{mekal}, without using the CTI corrector.
There is generally very good agreement between \textsc{mekal} and
\textsc{spex}. The main difference appears to be that fewer regions
require multi-temperature fits statistically using \textsc{spex},
leading to the abundances for a few outlying points to be changed. If
you compare the quality of the fits between the two models there is
virtually no difference.

There are many more differences between the results obtained using
\textsc{apec} and \textsc{mekal}. We wish to emphasise that the
increased scatter in Fig.~\ref{fig:codecompar} does not imply that
\textsc{apec} is worse than \textsc{spex}, but just that it is less
similar to \textsc{mekal}. The O metallicities are substantially
larger using \textsc{apec}, the Mg abundances larger, whilst Ni is
slightly smaller. Ne is poorly correlated, indicating that these
models disagree over the peak of the Fe-L emission. The models agree
over Si, S, Ar and Ca metallicities. There are some differences for
regions of high Fe abundance. In Centaurus this is where the gas is
the coolest and so Fe-L lines are more important, and multiple
temperature components more prevalent. The differences are also
slightly seen at high Si and S metallicities. The quality of the fits
produced by \textsc{apec} are generally better than the \textsc{mekal}
model.

\subsection{Metal determination}
With the low spectral resolution CCD data we are using here, we cannot
determine metallicities using the strengths of individual lines.
Spectral models are used to predict the spectrum, which is folded
through the instrument response, and compared to the real data. The
problem with this approach is if the spectral model is incomplete or
inaccurate, or the calibration of the instrument is uncertain, then
derived abundances can be wrong.

\begin{figure}
  \includegraphics[width=\columnwidth]{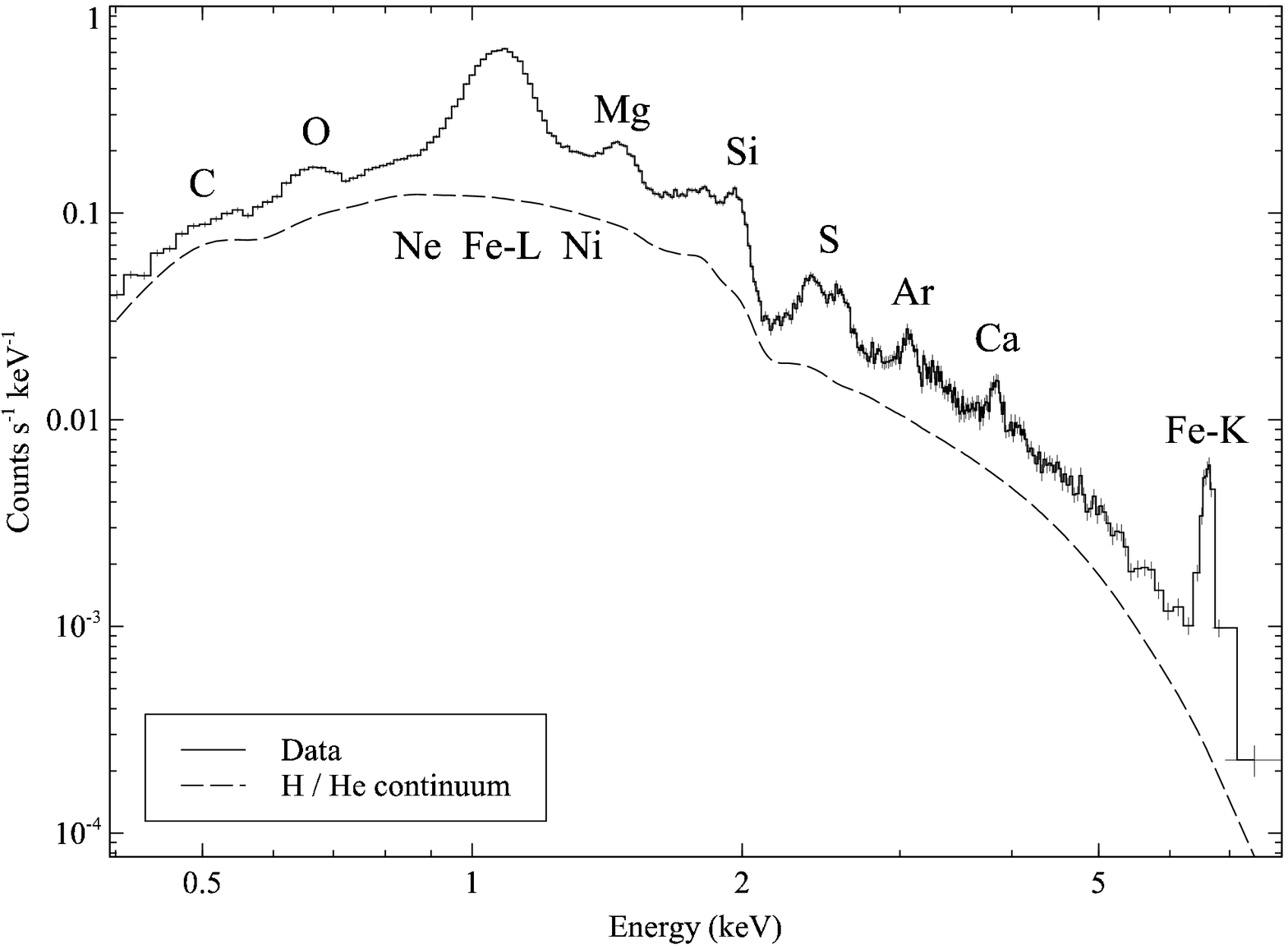}
  \caption{Example spectrum from annuli 3--6 of the western sector
    (see Fig.~\ref{fig:profiles}).  Also plotted is a two-component
    model fit where the abundances have been set to zero afterwards,
    except for H and He. Labelled are the positions of the strongest
    lines of each element.}
  \label{fig:examplespec}
\end{figure}

However if an element produces strong lines, or lines which are
separated in energy from other elements, then its metallicities are
less likely to suffer from systematics. We show in
Fig.~\ref{fig:examplespec} a spectrum from the western sector of the
cluster, with a radius of 2.6 to 12.6~kpc. The main line positions are
labelled.

It can be seen that the strongest line at this temperature range is
the Fe-L line complex. This is especially strong in the cooler region
of the cluster. Further out the Fe-K lines become more important. The
Fe-L lines can be very sensitive to the temperature of the gas, and so
it is important to include a sufficient range of the temperature
components in the spectral fitting to ensure the abundance is
accurate. There is some additional complexity as the Ne, Mg and Ni-L
lines are close in energy to Fe-L, and these can affect the Fe-L
abundance. The strength of the Fe lines means that the statistical
error on its metallicity is very low. We do use at least two spectral
codes in our analysis, however, and they show very similar trends.

The Si lines are the next strongest. Unfortunately there is another
potential systematic, which is that the Ir coating of the
\emph{Chandra} mirrors produce a steep drop in the effective area of
the at around 2.05~keV due to the Ir M edge (see spectrum). This means
that if the effective area is not calibrated sufficiently well, this
could affect the Si signal. \emph{XMM} shows a similar feature near
2~keV, due to an Au edge, affecting Si determination. However, the Si
abundances obtained with \emph{Chandra} and \emph{XMM} agree well (See
Section~\ref{sect:profiles}). Some of the weaker Ni and Mg lines
affect the Si lines.

S is a robustly determined element, with a small leakage of signal
from the neighbouring Si lines.

Ar and Ca should be accurately determined elements as their lines lie
by themselves in the spectrum. However their signal is fairly weak,
and the level of the continuum can have some effect on their values.

The O lines are quite strong. However, there has been some buildup of
a contaminant on the ACIS detectors (See Sanders et al 2004 for an
image of this material). Recent versions of the calibration include a
position-dependent correction for this material. This is still,
however, a difficult measurement to make due to the calibration
uncertainties at low energy. The derived O abundances do not match
between the \emph{Chandra} and \emph{XMM} data
(Section~\ref{sect:profiles}), but their trend appears in the same
direction.

At higher temperatures the Ni-K are robust metallicity determinants.
At these lower temperatures Ni is mixed in with the Fe-L complex and
Ne lines. Nevertheless there is reasonable between the measurements
made by the different spectral codes (Section~\ref{sect:speccodes}),
and between \emph{XMM} and \emph{Chandra}
(Section~\ref{sect:profiles}).

N and Al lines are weak, so we ignore these element, fixing them at
solar values in the spectral fits. C is fairly weak and lies at the
edge of our energy range, and so we also leave this abundance fixed
to a solar value.

\subsection{CTI correction}

\begin{figure}
  \includegraphics[width=\columnwidth]{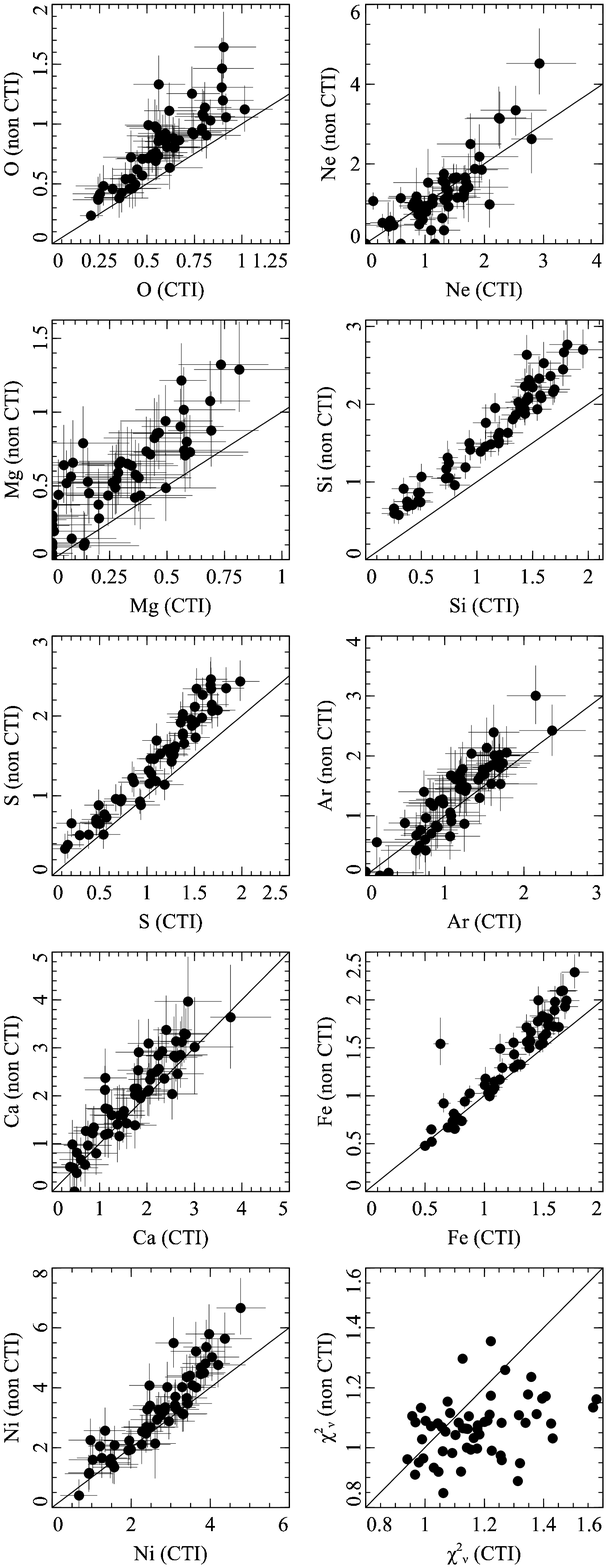}
  \caption{Comparison of abundances obtained using CTI and non CTI
    corrected data, relative to solar. Results shown are using the
    \textsc{spex} spectral model with one, two or three temperature
    components. Also shown are the reduced $\chi^2$ from the fits
    against the two datasets.}
  \label{fig:cticompar}
\end{figure}

It was unclear to us what effect the CTI correction would have on the
obtained abundances. We therefore examined the results using the CTI
corrector and using the standard analysis tools. We applied single,
double and triple temperature component fits with the \textsc{spex}
model to the data. In Fig.~\ref{fig:cticompar} is shown the abundances
obtained from the CTI corrected data for each region plotted against
those from data using the standard reduction software.

There are some differences between the results obtained from the two
different sets of data. For Fe, at low metallicities (and therefore at
comparatively high temperatures of $\sim 4$~keV), there is little
difference between the different datasets.  At higher metallicities,
use of the uncorrected data produces higher metallicities. The
uncorrected data show higher metallicities for each element,
especially at high abundances. More worryingly, some elements also
show offsets in their abundances. This difference is most readily
apparent in Si, S and O.

It is difficult to assess whether either of the two abundance
measurements are correct. We therefore compared the metallicities
obtained against those from \emph{XMM-Newton} data from the same
radius in the cluster (see Section~\ref{sect:profiles} for these
results). In the central regions, the relatively poor spatial
resolution of \emph{XMM} means that the small scale temperature
variations means abundance determinations are more difficult. In
addition the wide point spread function (PSF) moves strong emission
from the centre of the cluster to the outskirts. However we see good
agreement between the non-CTI corrected \emph{Chandra} and \emph{XMM}
results for the best determined elements, especially in the outskirts.
O is a notable exception.  The CTI corrected data show an offset in Si
relative to the \emph{XMM} results.

We will therefore present results using the standard data preparation
tools, and not those using the results of CTI correction. We note,
however, that the abundance trends we observe are not removed by the
CTI correction. In addition Si is the only element with strong
difference between the CTI corrected and standard results.

\begin{figure*}
  \includegraphics[width=\textwidth]{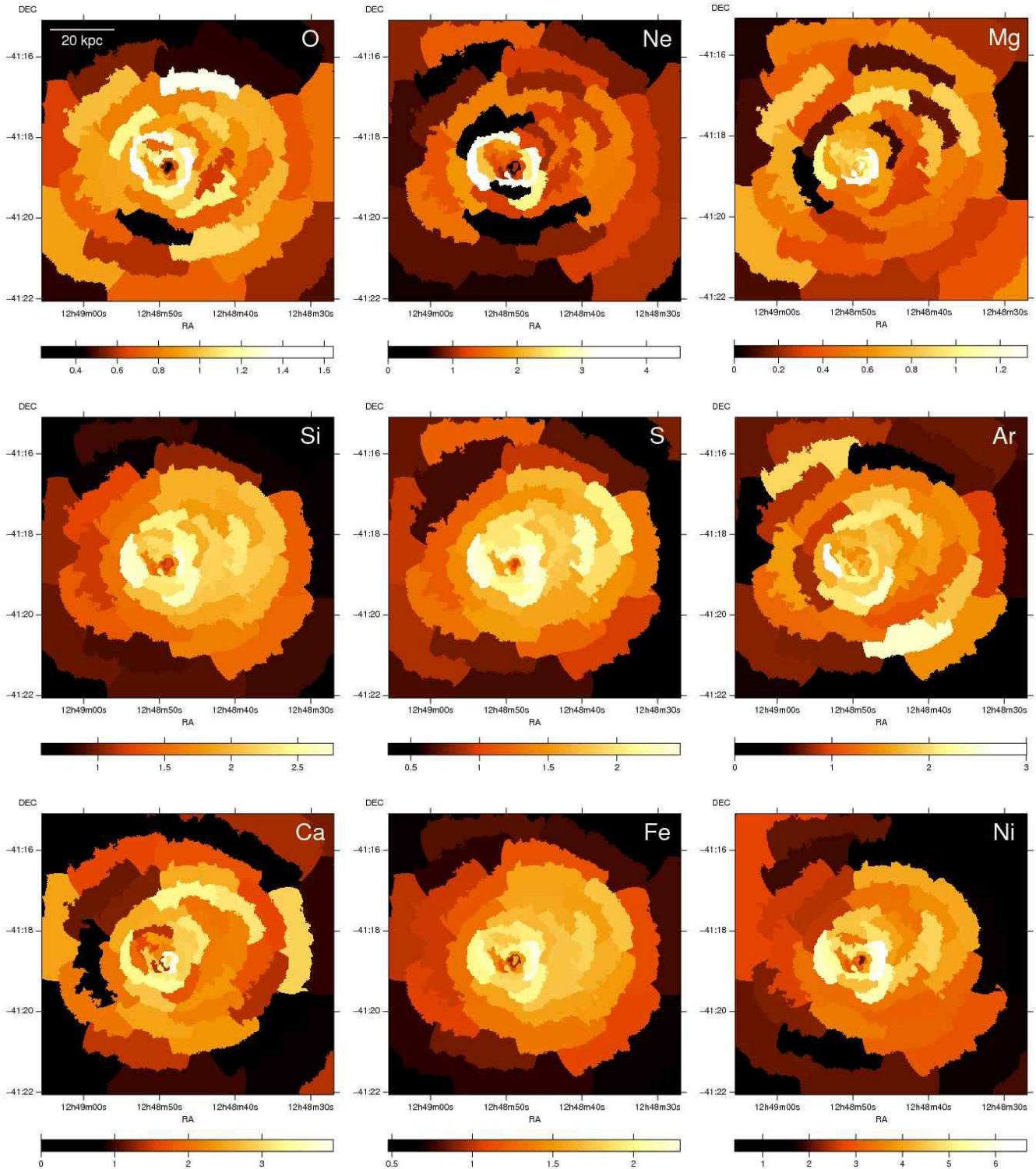}
  \caption{Abundance maps for each of the elements examined. The
    element maps were generated from spectra containing over 40\,000
    counts (signal to noise ratio of 200). These results were obtained
    using the \textsc{vspex} spectral model. The statistical errors on
    these quantities can be found using the scatter plots in
    Fig.~\ref{fig:abunscat}. The outer parts of these image appear to
    the bottom left of these plots, as these are the regions of lowest
    abundance.}
  \label{fig:abunmaps}
\end{figure*}

\begin{figure*}
  \includegraphics[width=0.8\textwidth]{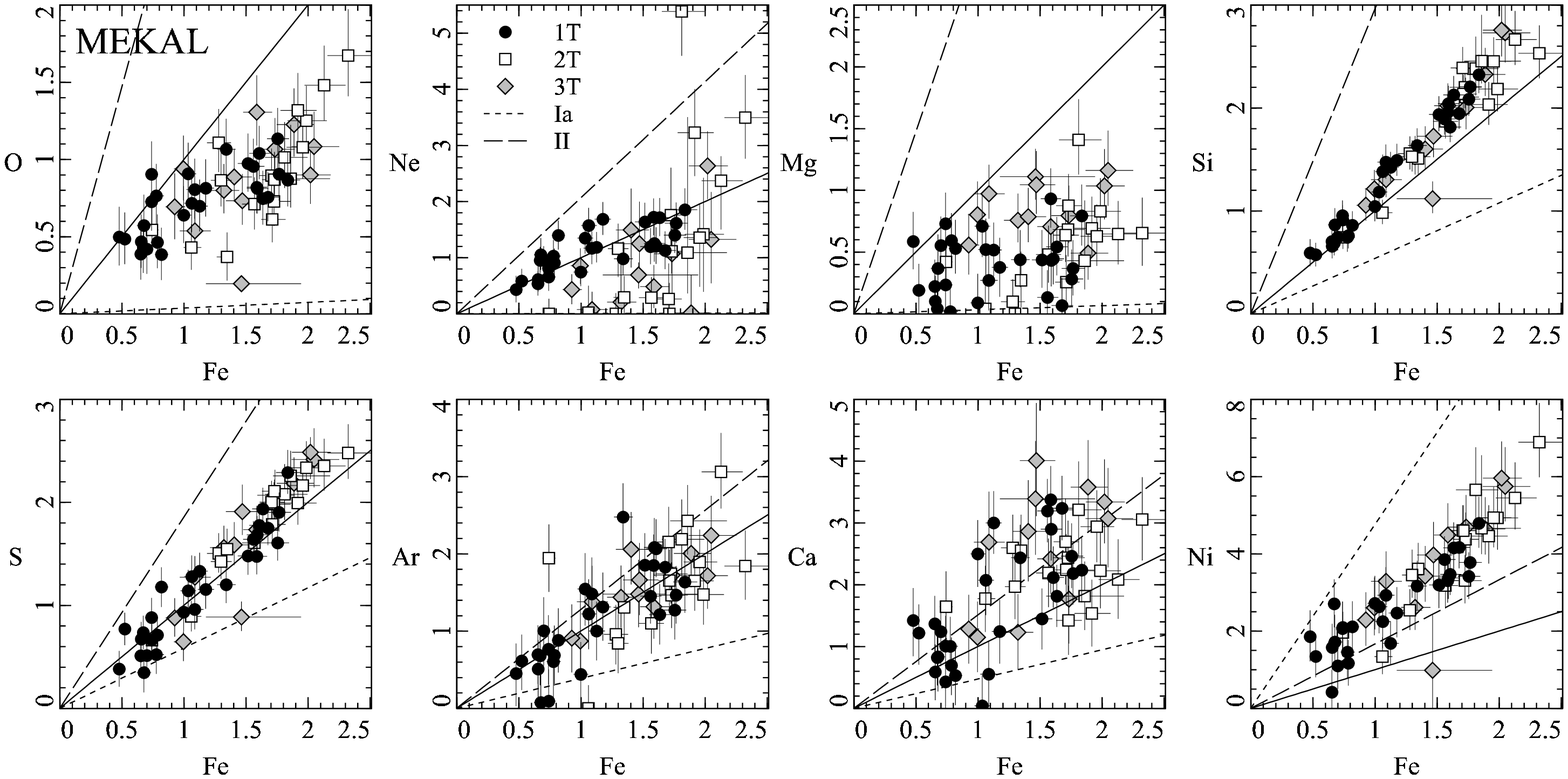} \\ \vspace{3mm}
  \includegraphics[width=0.8\textwidth]{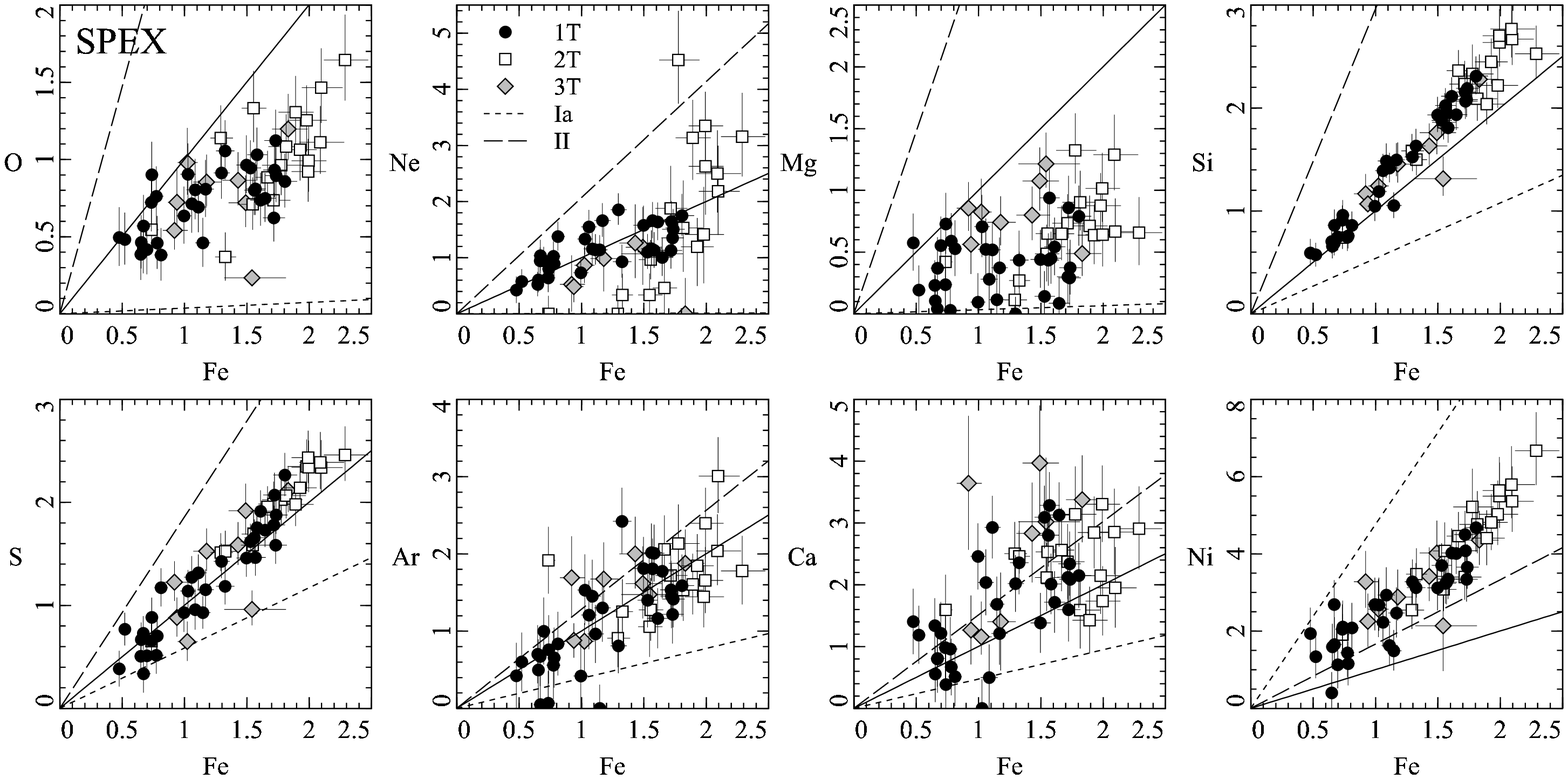} \\ \vspace{3mm}
  \includegraphics[width=0.8\textwidth]{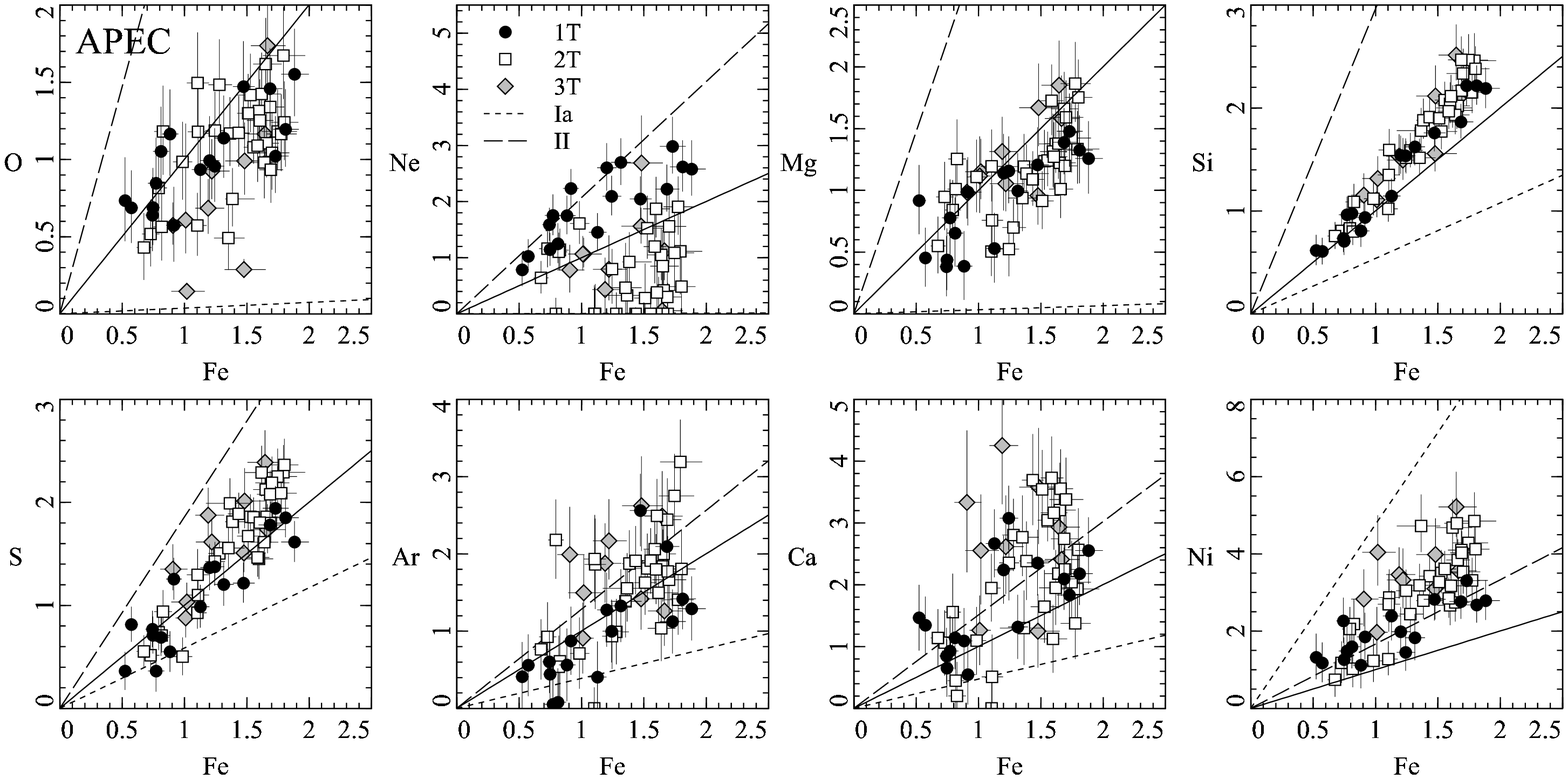}

  \caption{Abundance of individual elements of each spatial region
    from the 200 signal to noise ratio spectral fits, plotted against
    the iron abundance. The diagonal line shows solar abundance
    ratios.  White points show regions where two temperature model
    were required, black points show regions where single temperature
    models were sufficient, and grey points show where three
    components were required. Results using (top) \textsc{mekal},
    (centre) \textsc{spex}, and (bottom) \textsc{apec} plasma codes.
    Also plotted are the abundance ratios expected for the W7 Type~Ia
    model (Nomoto et al 1997b), and a Type~II model (Nomoto et al
    1997a).}
  \label{fig:abunscat}
\end{figure*}

\subsection{Maps}
\label{sect:maps}
In Fig.~\ref{fig:abunmaps} we map the abundance distributions in each
of the fitted elements, using the \textsc{spex} spectral model, with
one, two or three temperature components. In Fig.~\ref{fig:abunscat}
we also plot the values of the abundance measurements in each region
against the Fe abundance for the same region. On these plots are also
plotted the ratios of the abundances to Fe expected for the W7 Type~Ia
supernova model (Nomoto et al 1997b), and a Type~II model (Nomoto et
al 1997a). We examine the enrichment by different supernovae types in
Section~\ref{sect:snfits}.

The region where two temperature components are used is mainly within
80~arcsec radius of the core. Three temperature components are primarily
required within the central tail-like feature. It is interesting to
note that the third temperature in the inner regions is relatively hot
at around 6~keV, with the other two components at 0.7 and 1.5~keV.
This may suggest non-thermal or shocked emission is present in this
region.

\begin{figure}
  \centering
  \includegraphics[width=0.8\columnwidth]{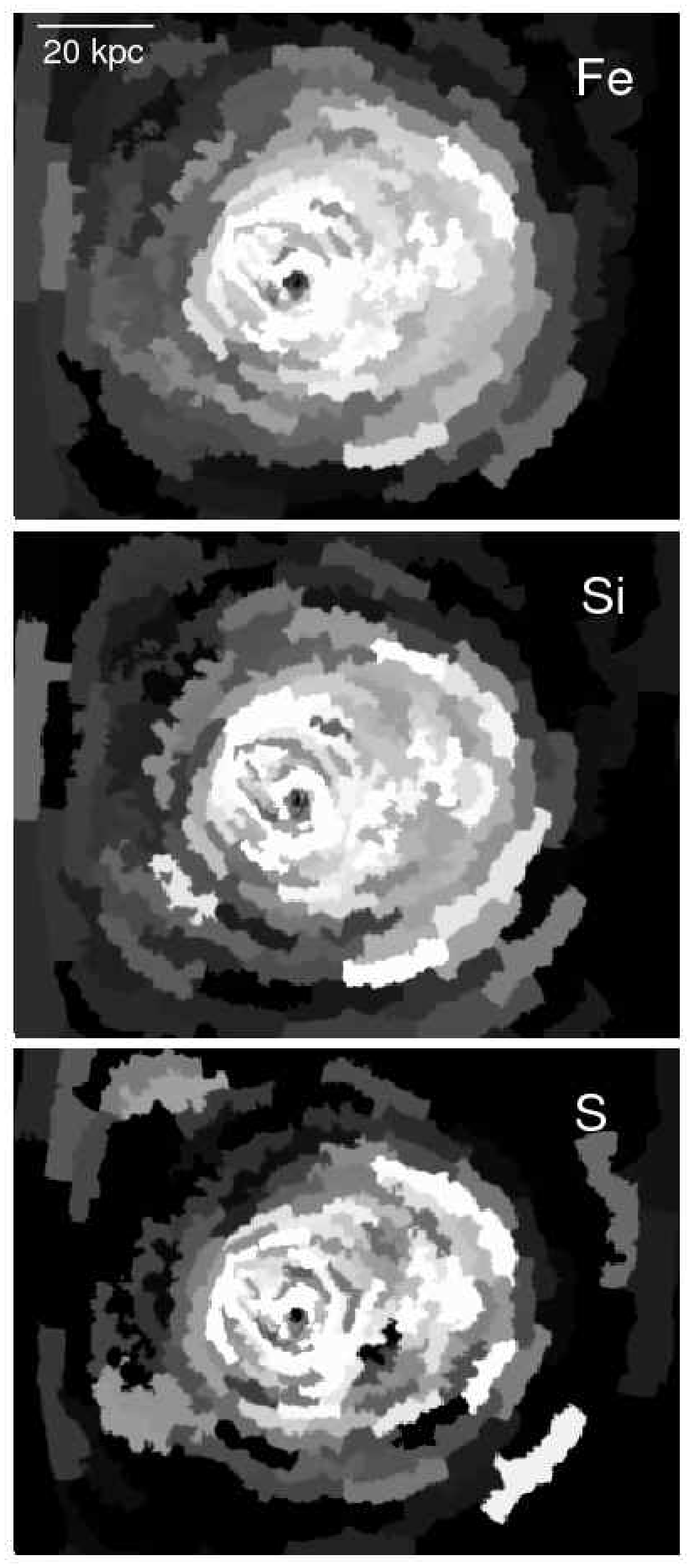}
  \caption{Higher spatial resolution abundance maps of Fe, Si, and S,
    generated using regions containing at least 10\,000 counts. Note
    the high metallicity plume-like feature embedded in the high
    metallicity extension to the west. There also appears to be high
    metallicity regions to the edge of the western extension.}
  \label{fig:abunmapshighres}
\end{figure}

\begin{figure}
  \centering
  \includegraphics[width=\columnwidth]{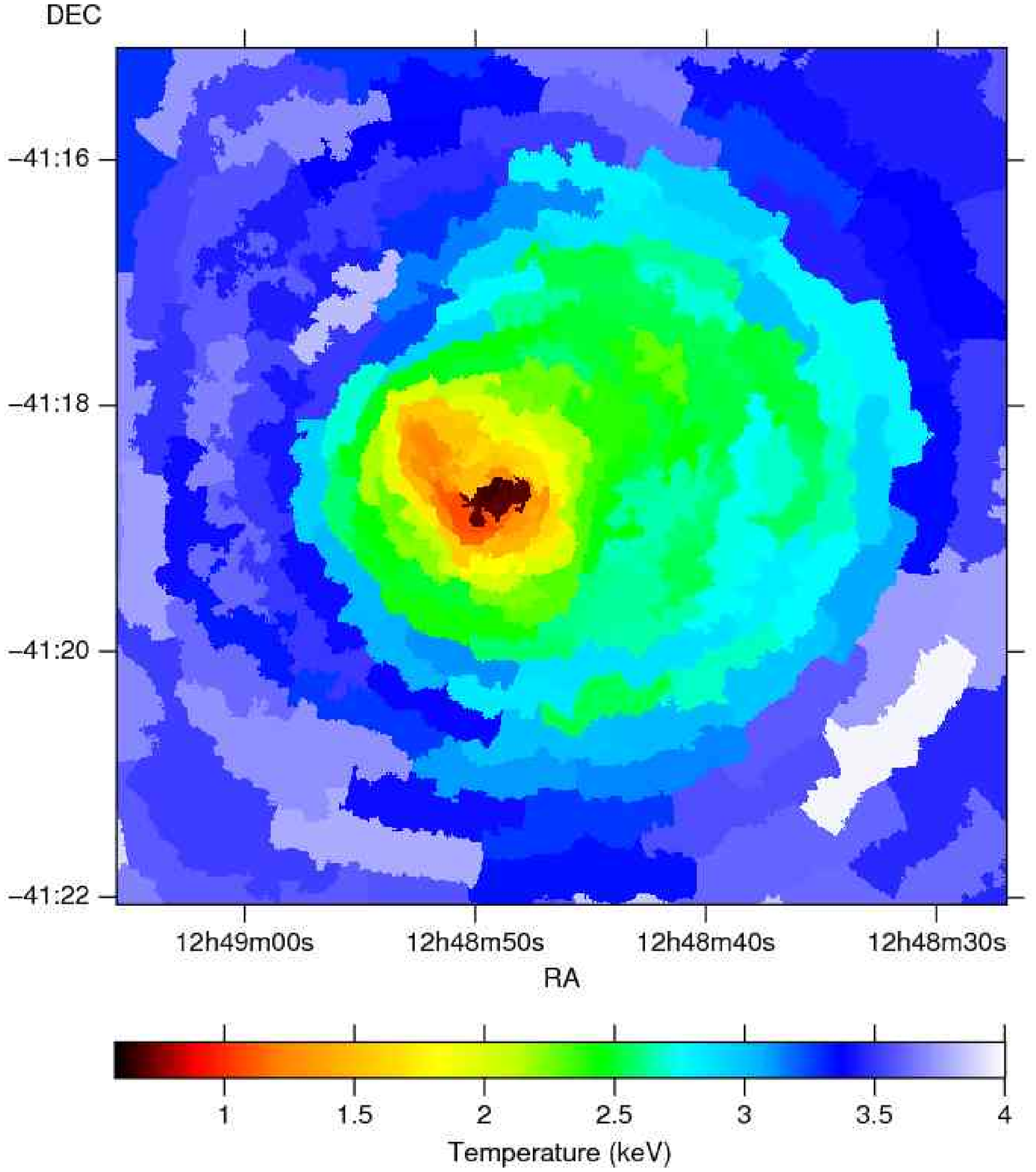}
  \caption{Emission weighted temperature map, using regions containing
    at least 10\,000 counts.}
  \label{fig:T}
\end{figure}

The best determined elements, including Fe, S and S, show extensions
to the west of the nucleus. To the west, the metallicities drops
rapidly at a radius of 30-40~kpc. This coincides with the rapid rise
in temperature (Fig.~\ref{fig:T}; Fabian et al 2005).

\begin{figure}
  \centering
  \includegraphics[width=\columnwidth]{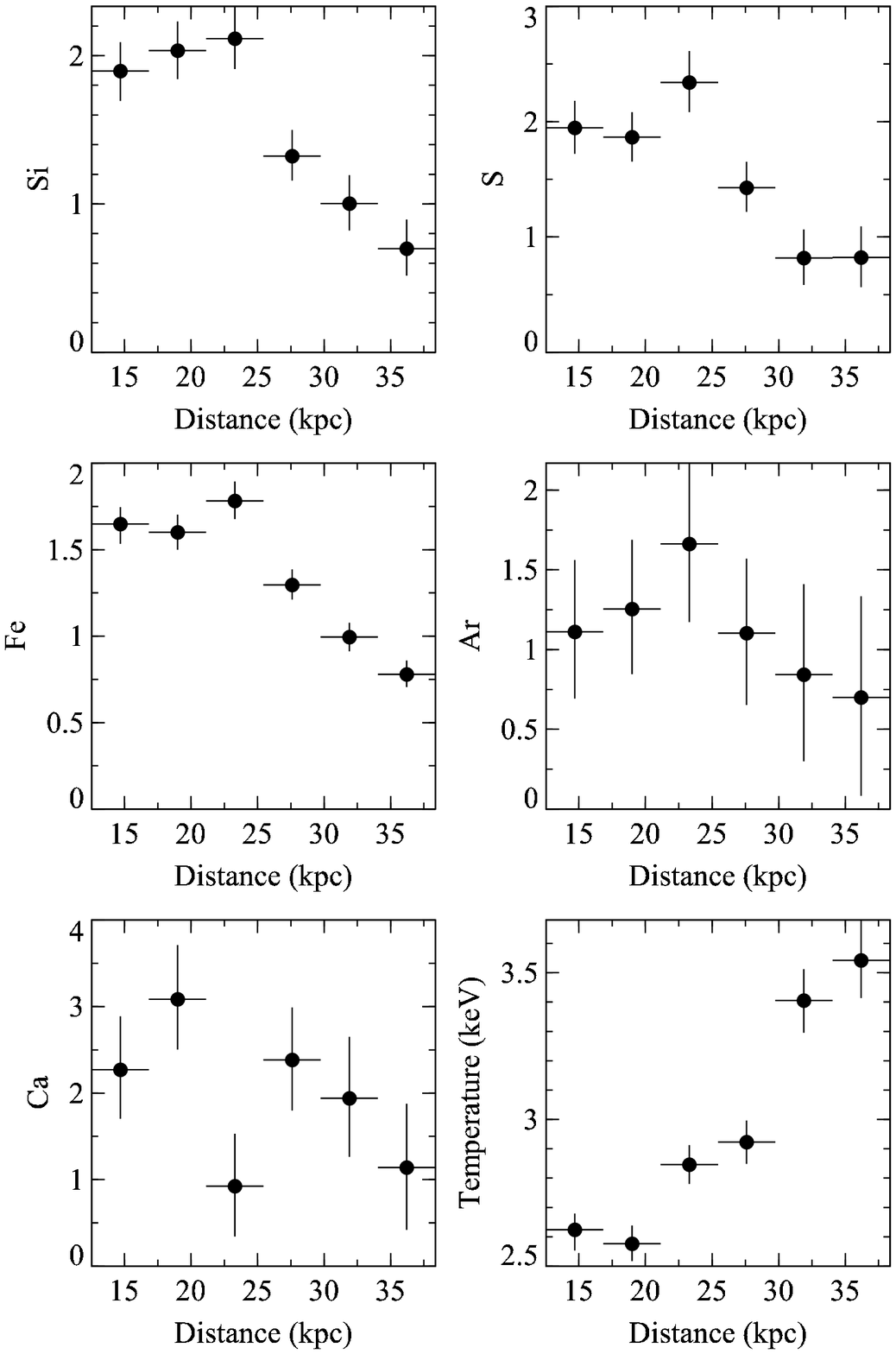}
  \caption{Profile across the edge of the high abundance region to the
    NW. There is some evidence for a spike in metallicity just inside
    the drop in some elements, include Fe, S and Si (third point from
    the left). Curiously there is a rise in temperature associated
    with the radius of the increase. The sectors are centred on the
    position $12^{h}48^{m}42.9^{s}$, $-41^\circ 18' \: 41''$, and are
    between angles 13.6 and 81.6 degrees northwards from the west.}
  \label{fig:dropprof}
\end{figure}

Their distributions appear quite smooth using these large regions.
However if we bin the data using a minimum signal to noise of 100,
then we observe quite a large amount of structure
(Fig.~\ref{fig:abunmapshighres}). This includes a ``plume like''
feature showing in Fe, Si and S around position
$12^{h}48^{m}38.0^{s}$, $-41^\circ 18' \: 27''$. Curiously the
elements often show enhancements at radii just before they drop, most
easily seen to the NW, at the edge of the high abundance region. These
appear similar in appearance to the high abundance shell found in the
Perseus cluster (Sanders, Fabian \& Dunn 2005). In more detail, this
spike can be seen in the profile in Fig.~\ref{fig:dropprof}.
Interestingly the temperature appears to rise across the edge at the
radius of the spike, and not after it.

\section{Profiles}
\label{sect:profiles}
To examine in detail the abundance of the elements, we have performed
an analysis using two 180 degree sectors to the east and west of the
cluster, as the maps of the cluster show a clear east/west anisotropy
in the central region.  In addition we performed spectral fits using
360 degree sectors from an \emph{XMM-Newton} observation of the
cluster, to examine the abundances at larger radii.

\begin{figure}
  \includegraphics[width=\columnwidth]{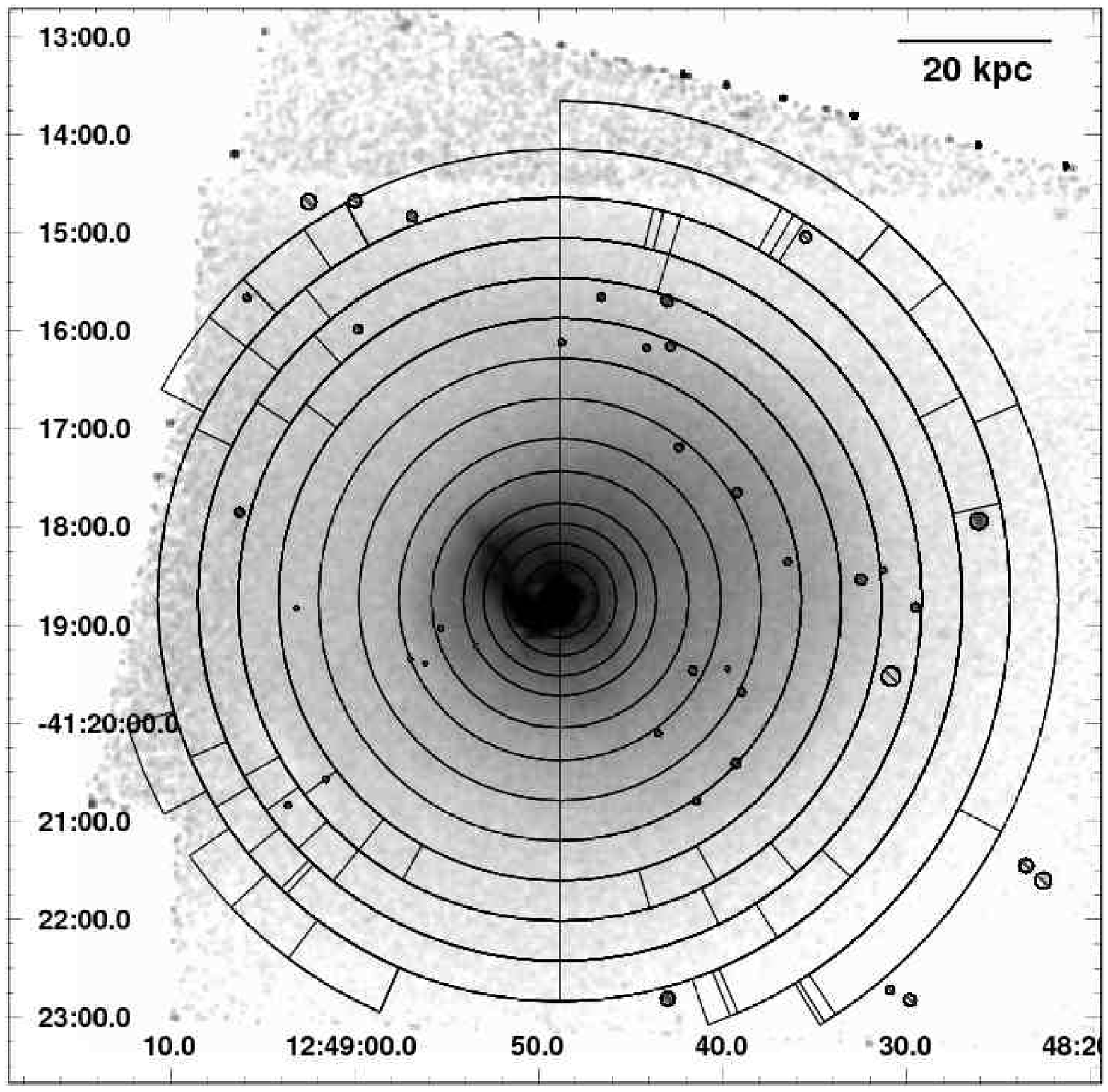}
  \caption{Regions used for the radial profiles in
    Section~\ref{sect:profiles}. Image is an exposure-corrected 0.4 to
    7~keV \emph{Chandra} image smoothed with a Gaussian of 1.5~arcsec.
    Point sources are marked with circles with lines across. The
    radial lines are starting and stopping angles of the regions for
    individual datasets which were taken with different roll angles.}
  \label{fig:regions}
\end{figure}

We show in Fig.~\ref{fig:regions} the \emph{Chandra} regions used for
this spectral analysis. As we analysed several different observations,
with different rolling angles and pointings, we used different angular
ranges for each dataset in the outer parts of the cluster.

\begin{figure*}
  \includegraphics[width=\textwidth]{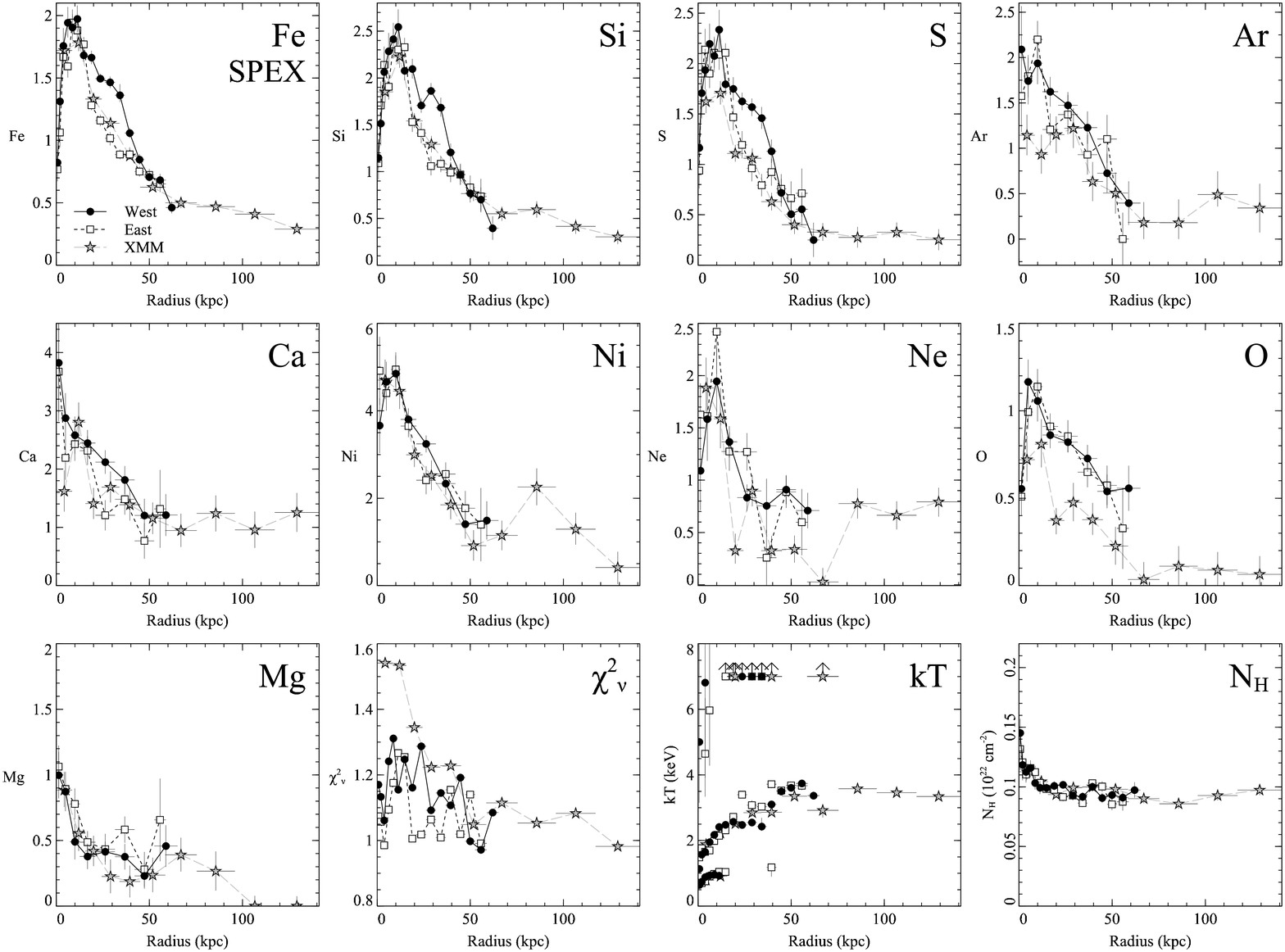}
  \caption{Projected abundance profiles for the west and east of the
    cluster, generated by fitting with the \textsc{spex} model. An
    F-test was used in each annulus to decide between 1, 2 and 3
    temperature component fits. Shown are the Fe, Si, S, Ar, Ca, Ni,
    Ne, O and Mg abundances relative to solar. Also shown are the gas
    temperature components and absorption fitted. If a temperature
    component is outside of the range shown an arrow at 7~keV is
    plotted.}
  \label{fig:profiles}
\end{figure*}

\begin{figure*}
  \includegraphics[width=\textwidth]{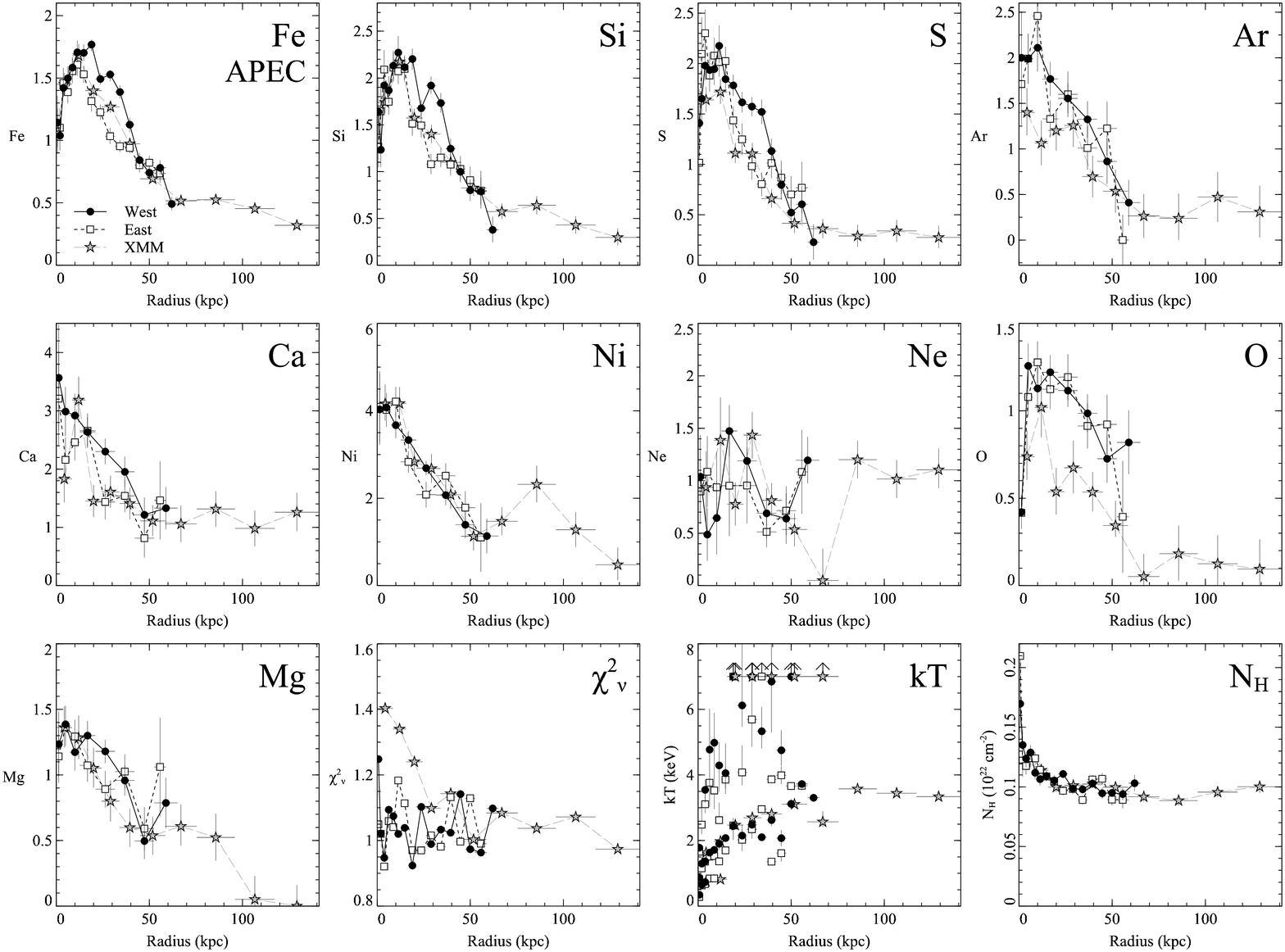}
  \caption{Same as Fig.~\ref{fig:profiles}, but using the
    \textsc{apec} spectral model.}
  \label{fig:profilesapec}
\end{figure*}

We used the same spectral fitting procedure as the maps for the
profiles. The derived abundances, temperatures and other properties
are all emission-weighted, projected results. In
Fig.~\ref{fig:profiles} are shown the \textsc{spex} model profiles for
each element for the east and west sectors, and the \emph{XMM}
results. Fig.~\ref{fig:profilesapec} contains similar profiles, but
using the \textsc{apec} model for comparison. We do not use the
\textsc{mekal} model here as we have already shown it produces very
similar results to \textsc{spex}.

The profiles show that each of the metals appears to be highly peaked.
As expected, in the inner regions the Fe, Si and S abundances are
enhanced to the west of the core compared to the east. This
enhancement appears to start at around 40~kpc radius. Within 15~kpc
there is no obvious difference between the two halves of the cluster.
These results agree with the two dimensional results in
Section~\ref{sect:maps}.

Curiously there is no obvious difference in the metallicities between
the two halves in Ar, Ni, Ne, O and Mg. There may be some enhancement
in Ca, however. These metals are harder to determine than Fe, Si and
S, but if the results are correct it suggests that the process causing
the east-west anisotropy enhances only Fe, Si, S and perhaps Ca. The
two-dimensional maps also agree with the profiles.

\begin{figure}
  \includegraphics[width=\columnwidth]{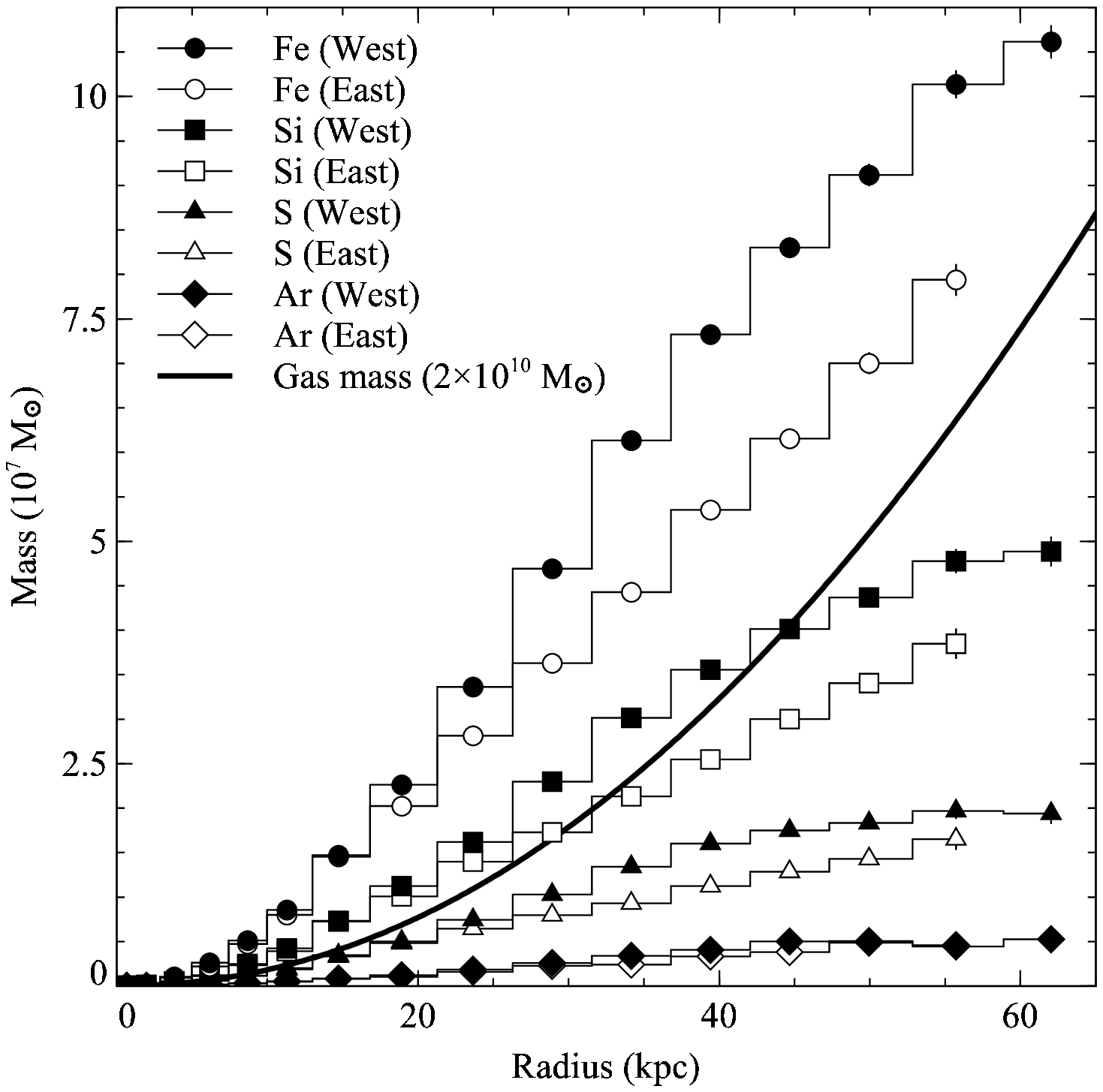}
  \caption{Cumulative mass profiles for the \emph{Chandra} results in
    the East and West sectors, for selected metals. Note that most of
    the metals show more mass to the west, except for Ar here. Also
    plotted is the cumulative gas mass (in units of $2\times
    10^{10}\Msun$), computed from the hydrogen density profile of
    Graham et al (2006).}
  \label{fig:cumlmass}
\end{figure}

This can be seen in a plot of the cumulative mass in different
elements as a function of radius (Fig.~\ref{fig:cumlmass}). These
results assume our projected \emph{Chandra} metallicities, but were
calculated with the deprojected density profile fit of Graham et al
(2006).  A base metallicity of $0.3\Zsun$ was subtracted from each
metallicity profile.  The density profile includes results from
\emph{Chandra}, \emph{XMM} and \emph{ROSAT}, going out to larger radii
than our \emph{Chandra} data alone. Also plotted is the cumulative gas
mass profile computed from the density profile.

The gradient in metallicity in the central region, for both the east
and western halves, shows a steep increase at a radius of about
50-60~kpc. Outside of this radius the abundance profile is relatively
flat. This picture is the same for each of the elements, except Mg
which is difficult to measure and differs greatly between the two
models.

\begin{figure}
  \includegraphics[width=\columnwidth]{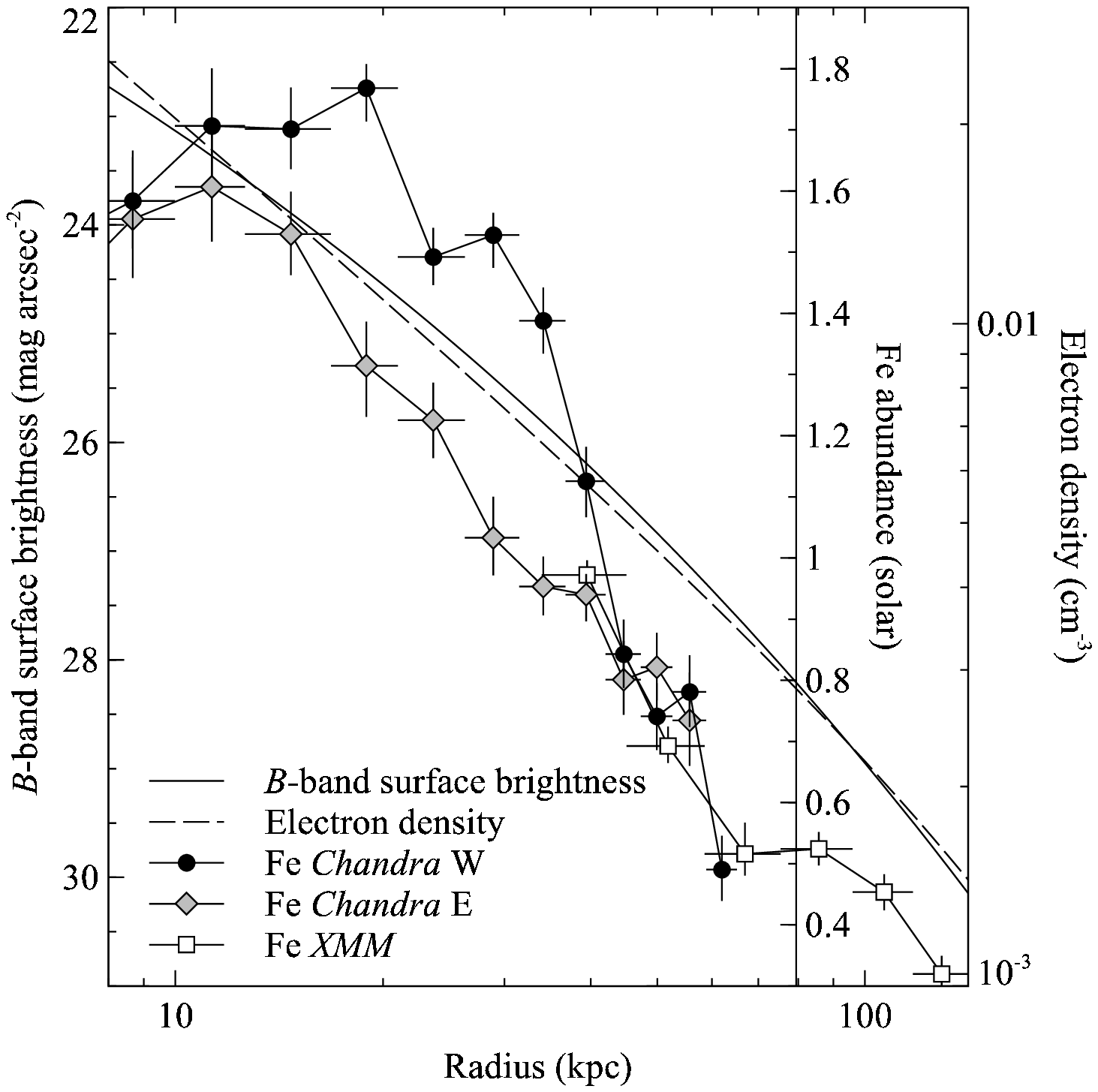}
  \caption{Comparison of the Fe metallicity, the X-ray electron
    density (Graham et al 2006), $B$-band surface brightness profile
    (Jerjen \& Dressler 1997). The $B$-band profile was computed using
    the effective radius (15.4~kpc) and the mean effective surface
    brightness.  We show the east/west \emph{Chandra} metallicity
    results in the inner region and the spatially averaged \emph{XMM}
    values at large radii.}
  \label{fig:optprofile}
\end{figure}

\begin{figure}
  \includegraphics[width=\columnwidth]{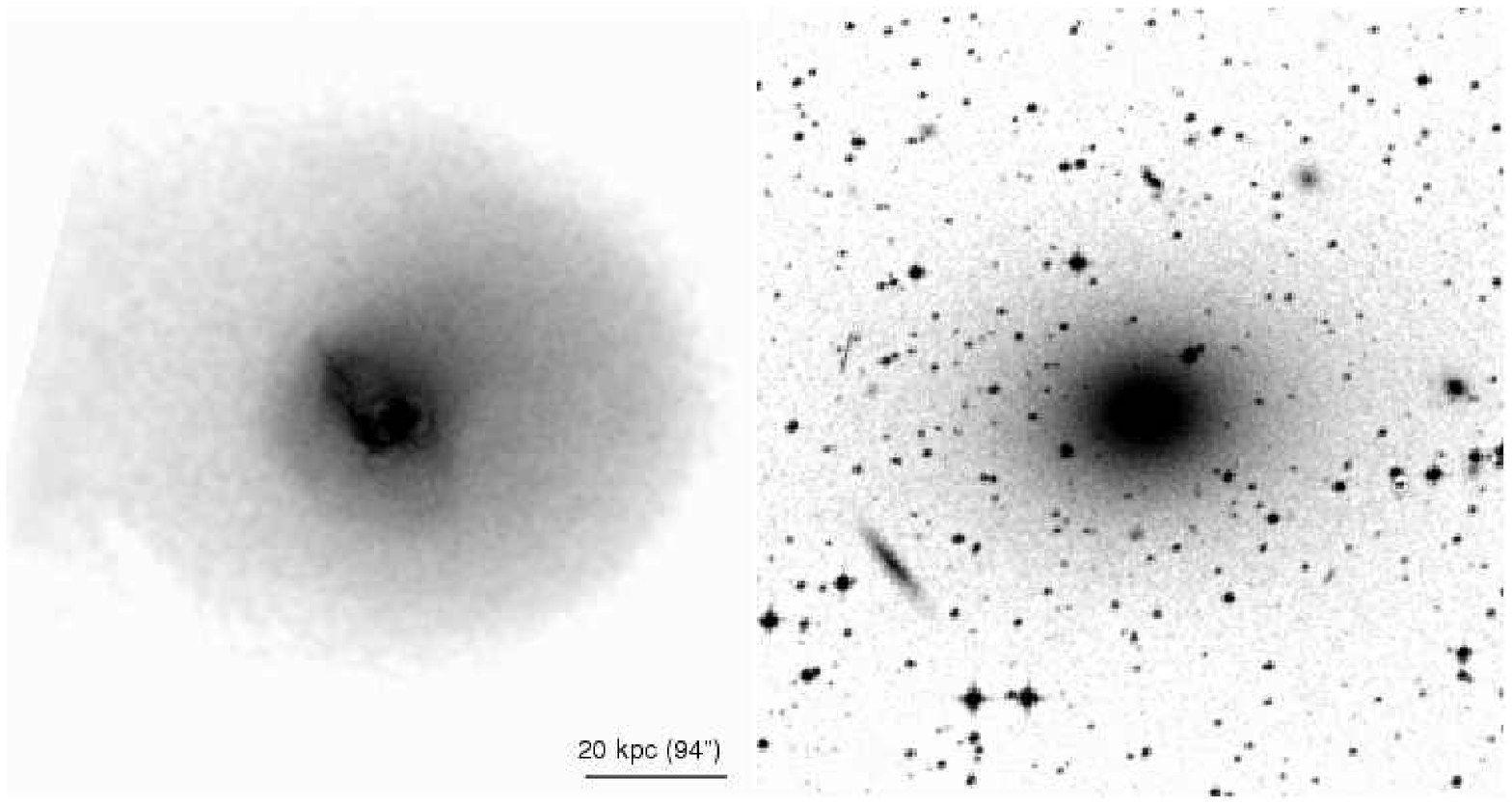}
  \caption{(Left) accumulatively smoothed (Sanders 2006) X-ray
    image in the 0.4-7~keV band with the point sources removed. The
    whole of the shaded region is metal rich.  (Right) Digitized Sky
    Survey image of the same area of sky.}
  \label{fig:xrayoptcompar}
\end{figure}

Fig.~\ref{fig:optprofile} shows the Fe metallicity profile compared
against the $B$-band surface brightness and electron density profiles.
Both the electron density and optical surface brightness profiles have
similar shapes. A comparison of X-ray emission (showing the metal rich
region) and DSS optical image of the cluster core
(Fig.~\ref{fig:xrayoptcompar}) shows that the metals lie over a large
region compared to majority of the optical light.

\subsection{Supernovae model fits}
\label{sect:snfits}
\begin{figure*}
  \includegraphics[width=0.9\textwidth]{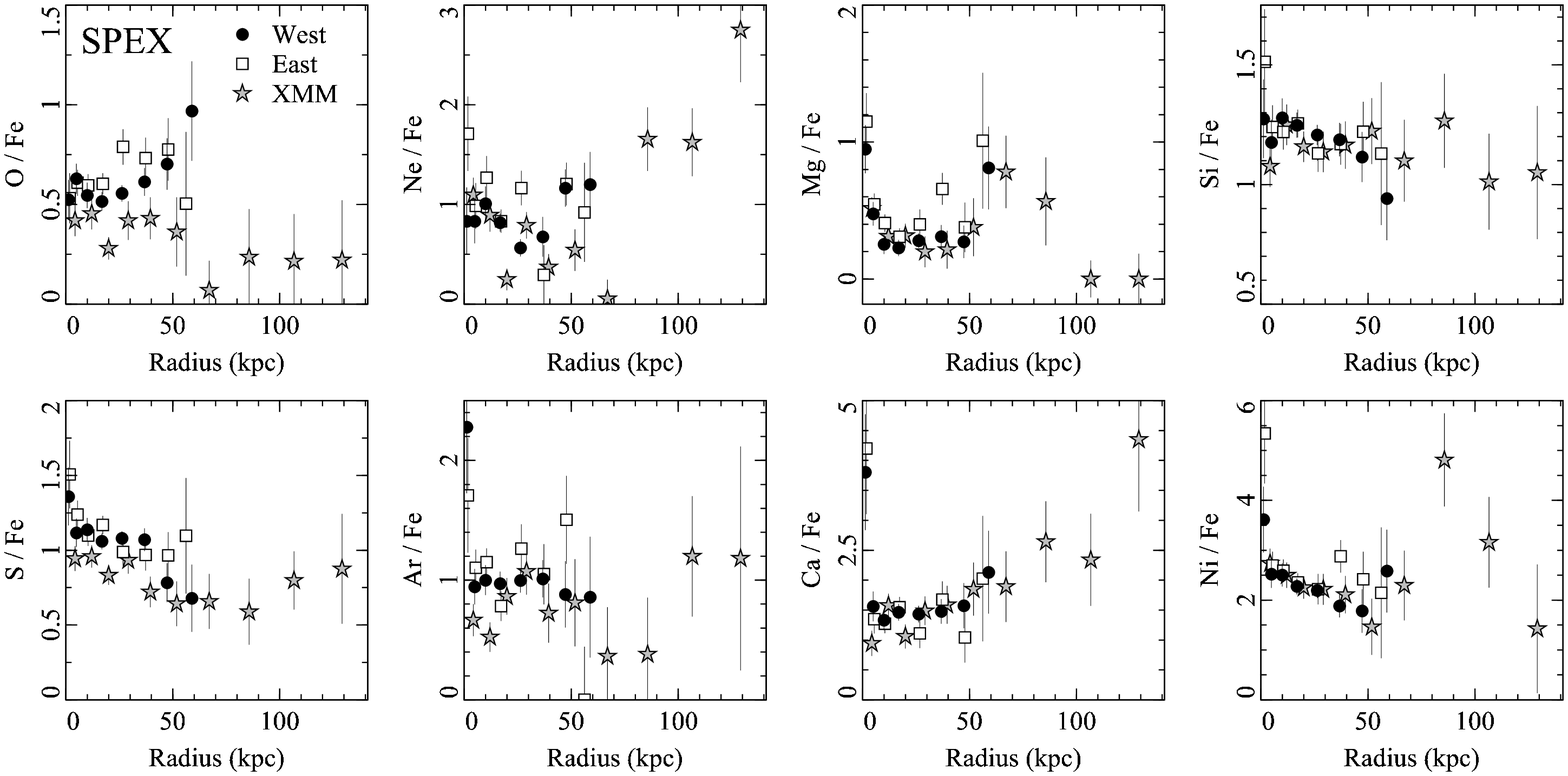} \\ \vspace{5mm}
  \includegraphics[width=0.9\textwidth]{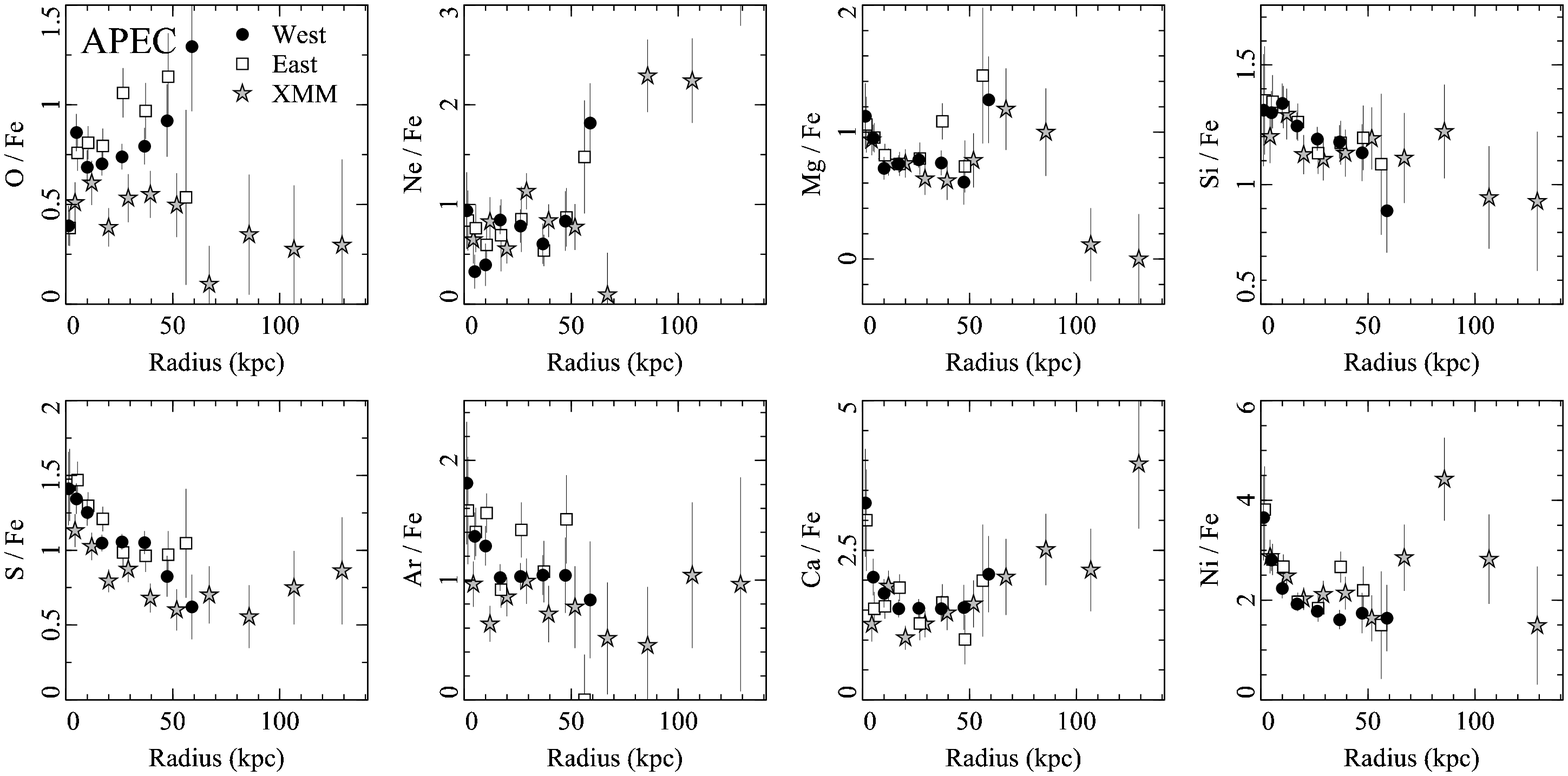}
  
  \caption{Profiles of the ratios of the solar relative metallicities
    of each element to Fe, using the \textsc{spex} (top) and
    \textsc{apec} (bottom) models.}
  \label{fig:ratioprofs}
\end{figure*}

To understand the the physical processes behind the enrichment, we
have examined the ratios of the abundances of the elements with
respect to Fe, to compare to predictions from supernovae yield models.
The ratios here are the ratios of the measurements in solar units.
Fig.~\ref{fig:ratioprofs} shows the ratios as a function of radius,
using the \textsc{spex} and \textsc{apec} models. The \emph{Chandra}
ratios have been binned as pairs to improve the signal to noise.

\begin{table}
  \caption{Ratios of elemental abundances in solar units relative to
    Fe for each SNe model. The
    Type Ia ratios were calculated for each element for the W7, W70,
    WDD1, WDD2 and WDD3 models in Nomoto et al (1997b). The Type II ratios were
    calculated from the results in Nomoto et al (1997a), integrating
    over an initial mass function of index 1.35 between 10 and 50
    solar masses.}
  \begin{tabular}{lllllll}
    \hline
    Element & W7    & W70   & WDD1  & WDD2  & WDD3  & Type II \\ \hline
    O       & 0.037 & 0.033 & 0.035 & 0.019 & 0.011 & 2.940 \\
    Ne      & 0.006 & 0.003 & 0.003 & 0.001 & 0.001 & 2.067 \\
    Mg      & 0.033 & 0.058 & 0.044 & 0.019 & 0.010 & 2.949 \\
    Si      & 0.538 & 0.467 & 1.688 & 1.013 & 0.632 & 2.961 \\
    S       & 0.585 & 0.597 & 1.972 & 1.199 & 0.747 & 1.857 \\
    Ar      & 0.387 & 0.463 & 1.424 & 0.895 & 0.554 & 1.281 \\
    Ca      & 0.473 & 0.718 & 2.138 & 1.399 & 0.868 & 1.511 \\
    Ni      & 4.758 & 3.637 & 1.650 & 1.398 & 1.572 & 1.663 \\
    \hline
  \end{tabular}
  \label{tab:modelratios}
\end{table}

In Table~\ref{tab:modelratios} we list the ratios calculated for
different supernova models. The Type Ia ratios (models W7, W70, WDD1,
WDD2 and WDD3) were calculated using Nomoto et al (1997b). The Type II
ratio was calculated from Nomoto et al (1997a), integrating over an
initial mass function of index 1.35 between 10 and 50 solar masses.

\begin{figure*}
  \includegraphics[width=0.9\textwidth]{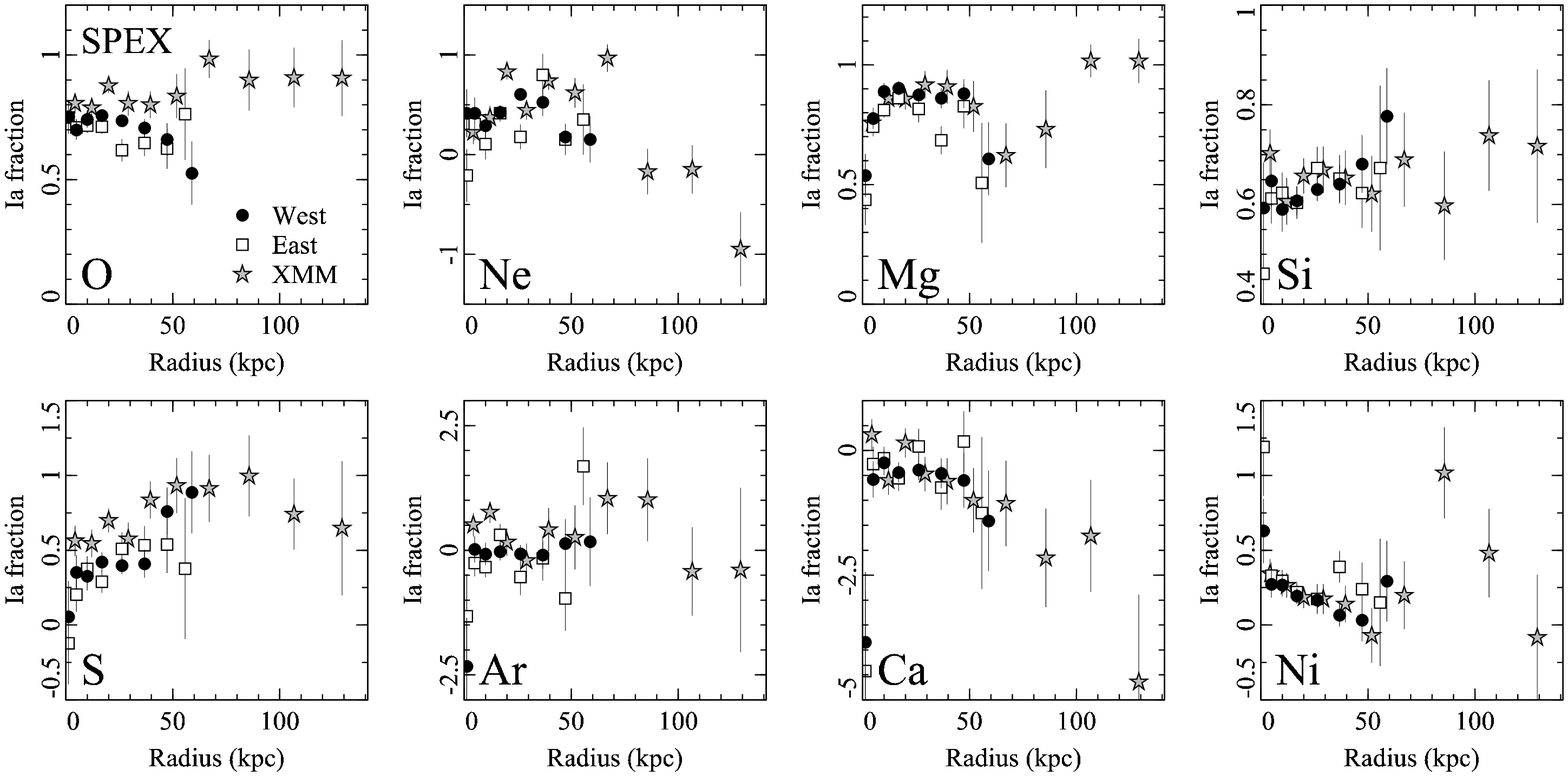} \\ \vspace{5mm}
  \includegraphics[width=0.9\textwidth]{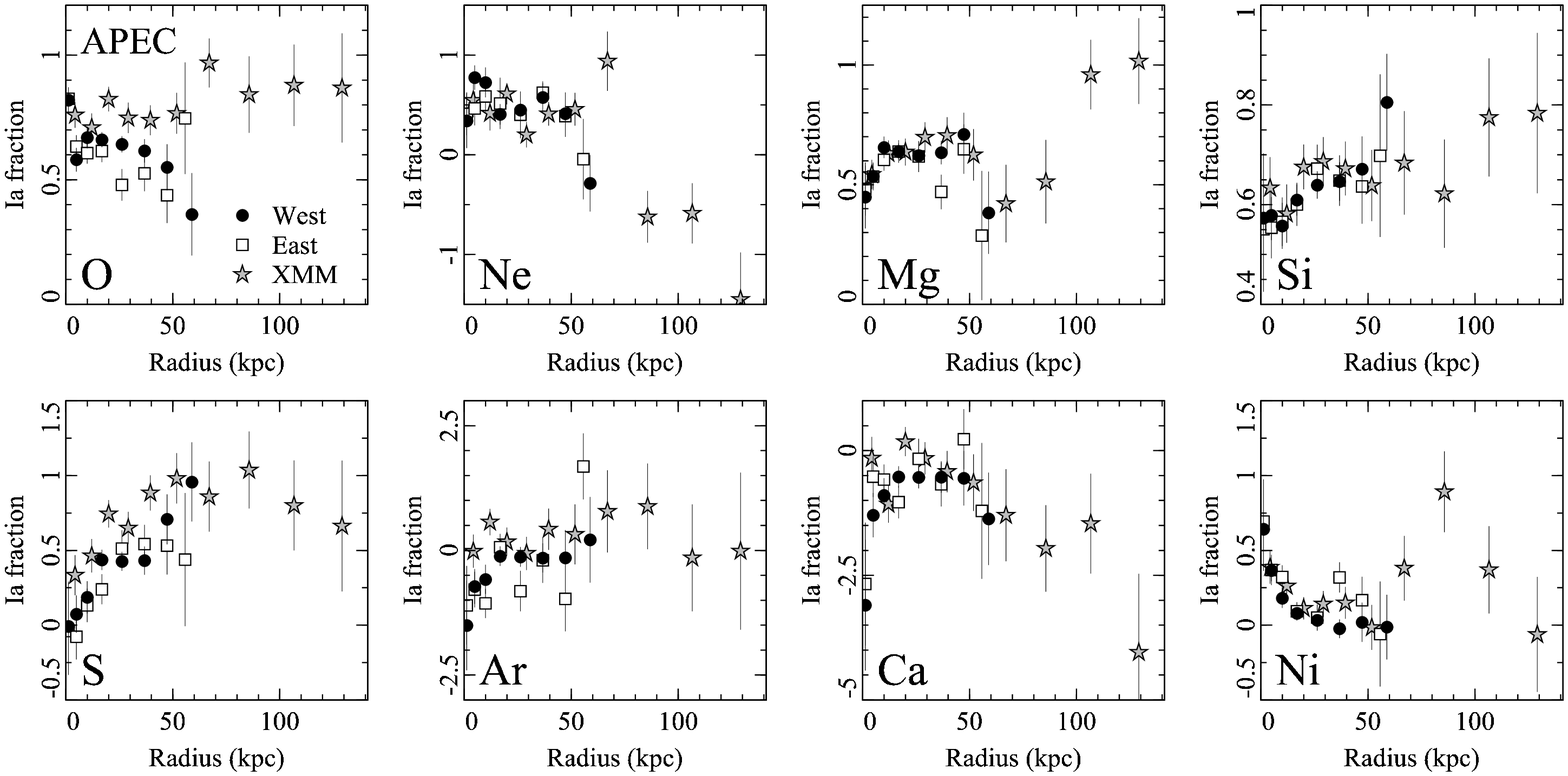}

  \caption{Calculated Type Ia fractions from the ratio of each element
    to Fe. Top panel shows \textsc{spex} and the bottom \textsc{apec}.
    These results assume the W7 Type Ia abundances, and Nomoto Type II
    results.}
  \label{fig:snfracs}
\end{figure*}

Taking an observed ratio for a particular element relative to Fe from
Fig.~\ref{fig:ratioprofs}, we can compute, assuming a Type Ia model
and Type II model, the ratio of Type Ia to II supernovae enrichment.
Shown in Fig.~\ref{fig:snfracs} are the fractions of the enrichment
due to Type~Ia supernovae, calculated using the ratio of each element
to Fe, as a function of radius. This is the result of solving the
equation
\begin{equation}
  \left( \frac{Z_\mathrm{X}}{Z_\mathrm{Fe}} \right) =
  f \left( \frac{Z_\mathrm{X}}{Z_\mathrm{Fe}} \right)_\mathrm{Ia} +
  (1-f) \left( \frac{Z_\mathrm{X}}{Z_\mathrm{Fe}} \right)_\mathrm{II},
\end{equation} 
for $f$, where $f$ is the Type Ia fraction, and $Z_\mathrm{X}$ is the
abundance of element X, relative to solar. The two terms on the right
are the predicted Ia and II ratios. Here we assume the yields of
Type~Ia supernovae using the W7 model, and the Nomoto et al (1997a)
Type II model.

The results show that for the best determined abundances, Si and S,
there appears to be an decreasing ratio of Type~Ia enrichment within
about 50~kpc radius. Ar and O ratios also indicates that this is case.
Ca, however, indicates the opposite, and so does Ni. Ni is a difficult
element to determine, however. Ne and Mg also appear to be dependent
on the spectral model used. We note that Werner et al (2006) and de
Plaa et al (2006) find Ca values which disagree with standard
nucleosynthesis results for the clusters 2A~0335+096 and S\'ersic
159-03, respectively.

\section{Discussion}
The \textsc{apec} results shown in Fig.~\ref{fig:snfracs} provide a
reasonably consistent picture from O, Ne, Mg, Si and S in which the
Type Ia fraction is about $0.7\pm0.1$, with a possible central drop
(within $10\kpc$) to 0.5. The remaining results (Ar, Ca and Ni) are
from weak or confused (Ni) lines. The \textsc{spex} results are less
conclusive but are also broadly consistent with this conclusion.  Such
a Type Ia fraction is very similar to that found both in the hot
interstellar medium of early-type galaxies (Humphrey \& Buote 2006)
and the Solar Neighbourhood.

Our results are therefore consistent with the expected products of
normal star formation. This conclusion contrasts with the usual model
(Matsushita et al 2003) discussed for the central iron excess in
clusters, which assumes that the excess iron is from Type Ia
supernovae alone. Such a Type-Ia-only picture is usually justified by
the assumption of no massive star formation in the central galaxy. and
thus no Type II supernovae, since the last major cluster merger or major
disruption. This may be the case for most clusters, but not for
central high abundance objects, like the Centaurus cluster, which have
presumably been relatively undisturbed (Graham et al 2006).

The centre of the Centaurus cluster shows good evidence from X-ray
spectra for at least some continued cooling of the innermost hot gas
($\sim 10\Msunpyr$; Sanders \& Fabian 2002) and has a web of
emission-line filaments and dust lanes (Crawford et al 2005). The
central galaxy, NCG\,4696, is a giant elliptical, with a de
Vaucouleurs profile and no evidence for an extended halo in
near-infrared $H$-band observations (Arnalte Mur et al 2006). The
optical spectral indicator $D_{4000}$ of the galaxy light is slightly
smaller than seen in field ellipticals, which may indicate some excess
blue light and thus ongoing star formation (Johnstone, Fabian \&
Nulsen 1987). The total mass of iron in the high abundance peak at
radii less than 40~kpc is about $1.4\times 10^8\Msun$
(Fig.~\ref{fig:cumlmass}).  If 30 per cent of this is from Type II
supernovae which release an average of $0.07\Msun$ per event (Renzini
2004), then $6\times 10^8$ supernovae are required. Each Type II
supernova requires a total of about $70\Msun$ of star formation so we
deduce a total star forming mass of $4\times 10^{10}\Msun$, assuming a
normal initial mass function.

A direct consequence of our work is the discovery of the Type II SN
products of a significant episode of star formation in the central
cluster galaxy, NGC\,4686, of the Centaurus cluster. We now explore
three scenarios for this enrichment: i) continuous star formation, ii)
sporadic activity over the past few Gyr and iii) a remnant of the
formation phase of the galaxy, which we assume to have been about
10~Gyr ago. For the last scenario iii) we include the possibility that
the massive stars giving rise to the Type II products occurred early
enough to enrich the lower mass stars which now dominate stellar mass
loss in the galaxy.

In order to obtain a limit on the current star formation rate in
NGC\,4696 (scenario i), we use the Mg$_2$ index, which is reduced by
the presence of a young stellar population. (The H$\beta$ index,
Trager et al (2005), would be better but is contaminated here by line
emission.)  It has been employed by Cardiel et al (1998) to reveal and
measure the star formation rate in other Brightest Cluster Galaxies.
Mg$_2$ has been measured in NGC\,4696 by Gorgas, Efstathiou \&
Arag\'on Salamanca (1990) and by Annibali et al (2006). There is no
simple standard with which to measure star formation departures from,
although Gorgas et al (1990) report a corrected offset of 0.01. We
adopt that value as an upper limit and note that metallicity can shift
the Mg$_2$ index to higher values (Cardiel et al 1998); as shown in
the present paper, the intracluster medium around NGC\,4696 is
particularly metal rich.  Using the model curves in Cardiel et al
(1998) we then obtain a fractional limit on the $V$ light due to star
formation of about 10 per cent and a star formation rate limit of
$6\Msunpyr$ using $M_B=-23$ for the whole galaxy (Jerjon \& Dressler
1997). This assumes continuous star formation over the past 5~Gyr,
giving a mass to $V$-light ratio of 2.4 (Gorgas et al 1990).  This
limit is just below the value of $8\Msunpyr$ derived from the above
star-forming mass assuming star formation spread over 5~Gyr. The
values become consistent if the timescale is increased to 8--10~Gyr.
Further factors to consider are that the star formation could have
been concentrated to the centre of the galaxy, where there is a large
dustlane and other dust features, and that it may have occurred in
bursts (scenario ii).

The alternative possibility is to assume that the metal enrichment is
a fossil of the formation of the main body of the galaxy (scenario
iii). If the Type II products have always been in the intracluster
medium, it requires that the central intracluster medium has been
largely undisturbed for up to 10~Gyr and constraints on diffusion of
metals in the gas (e.g. Graham et al 2006) become even tighter. The
Type II products may however have been incorporated in lower-mass
stars which are now releasing the material through stellar mass loss.
To investigate this possibility we use the work of Ciotti et al
(1991), updated with the values of Pelligrini \& Ciotti (2006).
Specifically we increase the stellar mass loss formula of Ciotti et al
(1991) by 20 per cent to $\dot M_{\star}= 3\times 10^{-11}L_B
t_{10}^{-1.3} \Msunpyr$, where time is measure in units of
$10t_{10}$~Gyr and $L_B$ is the total $B$-band luminosity of the
galaxy (we use $L_B=10^{11}\Lsun$). The Type Ia supernova rate is
taken as $R_\mathrm{SN}=1.8\times 10^{-13} L_B
t_{10}^{-1.1}$~yr$^{-1}$. The total stellar mass lost over the past
5~Gyr (8~Gyr) is then $2\times 10^{10}\Msun$ ($6.3\times
10^{10}\Msun$) with a mean (Type Ia) iron abundance of 1.35 (1.21). If
the stars were already enriched by Type II SN, we then have a
plausible explanation for our result in terms of stellar mass loss
over the past 8 or more Gyr.

In summary, only part of the effect we find can be explained by
continuous star formation. Stellar mass loss must occur and will have
contributed much of the gas within the inner 40 kpc radius, where the
total gas mass is about $6\times 10^{10}\Msun$
(Fig.~\ref{fig:cumlmass}). If the mass lost by stars was enriched by
Type II supernovae then, together with enrichment by Type Ia
supernovae, the gas accumulated over the past 8~Gyr would resemble the
medium around NGC\,4696. A robust conclusion is that the enriched
region has not suffered major disruption in the past 8~Gyr or more.

The evidence for Type II supernova products requires that either star
formation has continued at a rate of $\sim 5\Msunpyr$, which should be
testable with further optical spectroscopy, or they are a remnant of
the original major star formation episode of NGC\,4696. The core of
the Centaurus cluster must have been in a close heating -- cooling
balance for the past 8~Gyr or more. Any major cooling or heating
episode would have reduced the metallicity peak either by consuming it
in star formation or by diluting it if the gas was driven to larger
radii. 

\section*{Acknowledgements}
The authors are grateful to Glenn Morris for his help in making the
spectra from the \emph{XMM-Newton} observation. We thank Luca Ciotti
for advice on stellar mass loss and enrichment. ACF thanks the Royal
Society for support.

\clearpage

\end{document}